\documentclass[nofootinbib,preprint,superscriptaddress,aps,prc]{revtex4-2}

\usepackage[utf8]{inputenc}

\usepackage{amsmath}
\usepackage{mathrsfs}
\usepackage{amssymb}
\usepackage{bbm}
\usepackage{graphicx}
\usepackage{graphics}
\usepackage[small,compact]{titlesec}
\usepackage{xspace}
\usepackage{colordvi}
\usepackage{color}
\usepackage{units}
\usepackage{stackrel}
\usepackage{hyperref}
\hypersetup{
    colorlinks,
    linkcolor={red!50!black},
    citecolor={blue!50!black},
    urlcolor={blue!80!black}
}

\usepackage[capitalise]{cleveref}

\usepackage{slashed}
\usepackage{orcidlink}

\usepackage{ulem}

\renewcommand{\emph}[1]{\textit{#1}}

\newcommand{\Nh}{\hat{N}}
\newcommand{\Ah}{\hat{A}}

\newcommand{\vect}[1]{\vec{\mathbf{#1}}}
\newcommand{\vectS}[1]{\vec{\boldsymbol{#1}}}

\newcommand{\EFT}{$\mathrm{EFT}(\slashed{\pi})$\xspace}

\newcommand{\comment}[1]{}
\newcommand{\oneb}{\widetilde{\boldsymbol{1}}}
\newcommand{\Db}{\mathbf{D}}
\newcommand{\Dbb}{\overline{\mathbf{D}}}

\newcommand{\jjvHe}{{}^3\mathrm{He}}
\newcommand{\jjvH}{{}^3\mathrm{H}}

\newcommand{\GT}{\left<\mathbf{GT}\right>}
\newcommand{\F}{\left<\mathbf{F}\right>}

\newcommand{\G}{\mathcal{G}}
\newcommand{\Gt}{\widetilde{\mathcal{G}}}
\newcommand{\Gb}{\boldsymbol{\mathcal{G}}}
\newcommand{\Mb}{\boldsymbol{\mathcal{M}}}
\newcommand{\Yb}{\boldsymbol{\mathcal{Y}}}
\newcommand{\Gammab}{\boldsymbol{\Gamma}}
\newcommand{\Gbt}{\widetilde{\boldsymbol{\mathcal{G}}}}

\makeatletter

\newcommand{\Rmnum}[1]{\expandafter\@slowromancap\romannumeral #1@}
\newcommand{\vast}{\bBigg@{4}}
\newcommand{\Vast}{\bBigg@{5}}

\begin{document}

\title{Coulomb Corrections to Three-Nucleon Moments}

\author{Ha S. Nguyen\,\orcidlink{0009-0006-1121-8568}}
\email{ha.s.nguyen@duke.edu}
\affiliation{Department of Physics, Box 90305, Duke University, Durham, North Carolina 27708, USA}

\author{Jared Vanasse\,\orcidlink{0000-0001-5593-6971}}
\email{jvanass3@fitchburgstate.edu}
\affiliation{Fitchburg State University, Fitchburg, Massachusetts 01420, USA
}

\date{\today}

\begin{abstract}
The Helium-3 ($\jjvHe$) magnetic moment and Gamow-Teller (GT) matrix element in triton ($\jjvH$) $\beta$-decay are calculated in pionless effective field theory (\EFT) to next-to-leading order (NLO).  Coulomb corrections are included perturbatively to $\mathcal{O}(\alpha)$ in this framework and should naively be $\alpha M_n/p^*\!\!\sim\!8\%$ corrections, where $p^*\!\!\sim\!88.5$~MeV is related to the three-nucleon binding momentum.  Fitting the two-nucleon iso-vector magnetic current low-energy constant (LEC), $L_1$, to the $\jjvH$ magnetic moment and the two-nucleon iso-scalar magnetic current LEC, $L_2$, to the deuteron magnetic moment we find the NLO $\jjvHe$ magnetic moment in units of nuclear magnetons is -2.130 and the surprisingly small $\mathcal{O}(\alpha)$ correction is 0.00335, $\approx\!0.18\%$ of the LO \EFT prediction. The leading-order (LO) GT matrix element for $\jjvH$ $\beta$-decay is 0.9806 while again it has a surprisingly small $\mathcal{O}(\alpha)$ Coulomb correction of $-0.000740$, $\approx\!0.08\%$ of the LO \EFT prediction.  At NLO we calculate the GT matrix element of $\jjvH$ $\beta$-decay, including the $\mathcal{O}(\alpha)$ Coulomb correction, in terms of the two-nucleon axial current LEC $l_{1,A}$.  Fitting $l_{1,A}$ to the $\jjvH$ half-life we make a prediction for the proton-proton fusion reduced matrix element of $\Lambda(0)=2.776(331)$.  Finally, we attempt to explain the unusually small size of the $\mathcal{O}(\alpha)$ corrections by investigating the Wigner-SU(4) expansion of these observables.
\end{abstract}


\maketitle
\newpage

\section{Introduction}

An ideal benchmark for how nuclear systems respond to external currents is provided by few-nucleon systems, where analytical calculations for various models of nuclear interactions can be done systematically. In this work we focus on weak and electromagnetic currents in three-nucleon systems.  Weak leptonic interactions with nuclei proceed through the exchange of W and Z bosons.  However, at the relatively low energies relevant for nuclear systems, the massive bosons can be integrated out leaving contact interactions with leptons known as the axial or Gamow-Teller (GT) current,  and vector or Fermi (F) current.  These interactions are important for understanding proton-proton ($pp$) fusion as well as $\beta$-decay and neutrino nuclear interactions.  Electromagnetic currents involve photons that either couple magnetically or directly to the charge of the nuclei.  These currents can be investigated through electron scattering or Compton scattering off of nuclei.

Properties of few-nucleon systems have traditionally been calculated via potential models.  A modern approach to calculate properties of few nucleon systems is provided by effective field theories (EFTs)~\cite{Hammer:2019poc,Bedaque:2002mn,Beane:2000fx,vanKolck:1999mw}.  Unlike potential models, EFTs allow for the rigorous estimation of errors in theoretical calculations.  In EFTs all terms allowed by the underlying symmetries are included in the Lagrangian.  This leads to an infinite number of terms that are organized by the EFT power counting, giving a finite number of terms at any order in the power counting and a tractable and systematically improvable methodology.  An EFT power counting is typically organized in powers of $p/\Lambda_b$, where $p$ is the typical momentum scale, and $\Lambda_b$ is the break down scale.  As $p\to\Lambda_b$ an EFT is said to break down as new degrees of freedom (\emph{i.e.}, new particles) need to be included.

At low energies ($E<m_{\pi}^2/M_N$) even pions can be integrated out, leading to a theory of contact interactions between nuclei and external currents known as pionless EFT (\EFT).  This theory has been used to great success for two~\cite{Chen:1999tn,Chen:1999vd,Rupak:1999rk}, three~\cite{Bedaque:1998mb,Bedaque:1999ve,Gabbiani:1999yv,Bedaque:2002yg,Griesshammer:2004pe,Vanasse:2013sda,Rupak:2001ci,Koenig:2011lmm,Konig:2013cia,Vanasse:2014kxa,Konig:2014ufa,Konig:2016iny}, and $A>3$ systems~\cite{Platter:2004zs,Stetcu:2006ey,Kirscher:2010dgl,Kirscher:2011uc,Bansal:2017pwn,Contessi:2017rww,Kirscher:2018dwo,Konig:2019xxk,Schafer:2022hzo,Bagnarol:2023crb}.  It has also been used to include external currents, looking at the charge and magnetic radii of two-~\cite{Chen:1999tn} and three-nucleon systems as well as magnetic moments~\cite{De-Leon:2020glu,Vanasse:2017kgh,Vanasse:2015fph} and magnetic polarizabilities~\cite{Kirscher:2017fqc}.  \EFT was also used to investigate interactions with neutrinos and dark matter, such as $pp$ fusion~\cite{Butler:2001jj,Kong:2000px}, triton ($\jjvH$) $\beta$-decay, two~nucleon neutrino scattering~\cite{Butler:2000zp,Butler:1999sv}, and two- and three-nucleon dark matter scattering~\cite{Richardson:2021liq}.  Previous calculations including external currents in the proton-deuteron ($pd$) system either ignored Coulomb interactions~\cite{Vanasse:2015fph,Vanasse:2017kgh} or included Coulomb interactions~\cite{De-Leon:2016wyu} as in Ref.~\cite{Vanasse:2014kxa} in which Coulomb is neither treated strictly perturbatively or non-perturbatively.  In the three-nucleon system Coulomb corrections naively scale as $\alpha M_N/p^*\!\!\sim\! 8\%$ where $p^*\!\sim\!\sqrt{(4/3)M_N|E_B+\gamma_t^2/M_N|}\!\sim\!88.5$~MeV, with $E_B$ and $\gamma_t$ the $\jjvH$ binding energy and deuteron binding momentum respectively.  This work follows the approach of Ref.~\cite{Konig:2015aka} in which Coulomb corrections were included strictly perturbatively in the $pd$ system to calculate the binding energy difference between $\jjvH$ and Helium-3 ($\jjvHe$).  However, unlike Ref.~\cite{Konig:2015aka}, which expanded the ${}^1S_0$ $N\!N$ channel about the unitary limit, we expand about the physical virtual bound state pole in the ${}^1S_0$ channel.  In this work we calculate the $\mathcal{O}(\alpha)$ Coulomb corrections to the GT and F matrix elements for $\jjvH$ $\beta$-decay and the $\jjvHe$ magnetic moment ($\mu^{\jjvHe}$).

An additional symmetry of nuclear systems is Wigner-SU(4) symmetry in which spin and isospin states of nucleons are combined into a single four dimensional object ($p \uparrow, p \downarrow, n \uparrow, n \downarrow$)~\cite{Wigner:1936dx}.  Wigner-SU(4) symmetry means the Lagrangian is invariant under arbitrary rotations in the SU(4) group for this space.  In the physical world Wigner-SU(4) symmetry is broken. However, its breaking can be treated perturbatively in the three-nucleon ~\cite{Lin:2024bor,Lin:2022yaf,Vanasse:2016umz} and four-nucleon~\cite{Chen:2004rq} systems.  Recently using the No-Core Shell Model (NCSM) with chiral-EFT based potentials Refs.~\cite{LiMuli:2025zro,Dang:2026xbw} showed that nuclear wavefunctions, up to $A=8$, are largely described by a single irreducible representation of the Wigner-SU(4) group.  Ref.~\cite{LiMuli:2025zro} used this observation to explain the observed pattern of suppressed $\beta$-decays in $A=3-8$ nuclear systems, while Ref.~\cite{Dang:2026xbw} suggests this observation could be useful in further refining the basis states used in NCSM calculations.  This paper considers the consequences of Wigner-SU(4) symmetry for the $\mathcal{O}(\alpha)$ Coulomb corrections to $\mu^{\jjvHe}$ and the GT and F matrix elements of $\jjvH$ $\beta$-decay.  We demonstrate that Coulomb corrections are additionally suppressed due to Wigner-SU(4) symmetry and are thus smaller than their naive estimate of $\alpha M_N/p^*\sim 8\%$. 

This paper is organized as follows.  Section~\ref{sec:Lag} gives the Lagrangian to NLO in \EFT including external currents.  In Sec.~\ref{sec:2N} the two nucleon system including perturbative Coulomb corrections is considered, while Sec.~\ref{sec:3N} does the same for the three-nucleon system.  Section ~\ref{sec:form} discusses the three-nucleon magnetic, charge, GT, and F form factors and Sec.~\ref{sec:wig} gives their Wigner-SU(4) expansions.  Finally, in Sec.~\ref{sec:results} we discuss our results and give conclusions in Sec.~\ref{sec:conclusion}.

\section{\label{sec:Lag}Lagrangian}

Using the dibaryon formalism the two-nucleon Lagrangian in \EFT up to and including NLO terms is given by
\begin{align}
&\mathcal{L}_2=\Nh^\dagger\left(iD_0+\frac{\mathbf{D}^2}{2M_N}\right)\Nh-\hat{t}_i^\dagger\left[\Delta_t-c_{0t}\left(iD_0+\frac{\Db^2}{4M_N}+\frac{\gamma_t^2}{M_N}\right)\right]\hat{t}_i\\\nonumber
&-\hat{s}_a^\dagger\left[\Delta_s-c_{0s}\left(iD_0+\frac{\Db^2}{4M_N}+\frac{\gamma_s^2}{M_N}\right)\right]\hat{s}_a+y\left[\hat{t}_i^\dagger\Nh^T P_i\Nh+\hat{s}_a^\dagger\Nh^T \Bar{P}_a\Nh+\mathrm{H.c.}\right].
\end{align}
Fields $\Nh$, $\hat{t}_i$, and $\hat{s}_a$ represent the nucleon, spin-triplet iso-singlet dibaryon, and the spin-singlet iso-triplet dibaryon respectively. $P_i=\frac{1}{\sqrt{8}}\sigma_2\sigma_i\tau_2$ ($\Bar{P}_a=\frac{1}{\sqrt{8}}\sigma_2\tau_2\tau_a$) projects out the spin-triplet (spin-singlet) combination of nuclei, where $i,a=1,2,3$.  The subscript on the Lagrangian denotes the number of nucleons involved and a subsequent value in the subscript of 0 or 1 denotes a specific \EFT order, LO and NLO, respectively.  The covariant derivative is defined by
\begin{equation}
    D_\mu\Nh=\partial_\mu\Nh+ie\mathbf{Q}\Ah_\mu\Nh,
\end{equation}
where the charge operator $\mathbf{Q}$ is $\frac{1+\tau_3}{2}$, $1$, and $1+T_3$ for the nucleon, spin-triplet dibaryon, and spin-singlet dibaryon respectively. $T_3$ gives the $z$ component of an iso-triplet state.  Parameters in the Lagrangian are fit using the $Z$-parametrization in which the bound (virtual bound) state pole of the ${}^{3}S_1$ ($^1S_0$) channel is fit at LO and its residue at NLO yielding~\cite{Phillips:1999hh,Griesshammer:2004pe}
\begin{equation}
    y^2=\frac{4\pi}{M_N}\quad,\quad \Delta_{\{t,s\}}=\gamma_{\{t,s\}}-\mu\quad,\quad c_{0\{t,s\}}=\frac{M_N}{2\gamma_{\{t,s\}}}(Z_{\{t,s\}}-1).
\end{equation}
$\gamma_t=45.7025$~MeV ($\gamma_s=-7.890$~MeV) is the ${}^3S_1$ bound state (${}^1S_0$ virtual bound state) pole momentum and $Z_t=1.6908$ ($Z_s=0.9015$)  its residue.  $\mu$ is a scale from using dimensional regularization with power divergence subtraction~\cite{Kaplan:1998tg,Kaplan:1998we}.

At LO a three-body force is required for renormalization group (RG) invariance and at NLO it receives a correction to avoid refitting~\cite{Bedaque:1998kg,Bedaque:1998km,Bedaque:1999ve}.  Defining the three-nucleon field $\hat{\psi}$, the Lagrangian for the LO three-body force and its NLO correction can be written as
\begin{align} \mathcal{L}_{3}=\hat{\psi}^\dagger\Omega\hat{\psi}+\sum_{n=0}^1\omega_0^{(n)}\left[\hat{\psi}^\dagger\sigma_i\Nh\hat{t}_i-\hat{\psi}^\dagger\tau_a\Nh\hat{s}_a+\mathrm{H.c.}\right].
\end{align}
This Lagrangian can be matched to a Lagrangian solely in terms of nucleon and dibaryon fields (See Ref~\cite{Vanasse:2015fph} for details).  In this work we fit the three-body force to the triton binding energy, $E_B=-8.481798$~MeV.  For details of how this is done see Ref.~\cite{Vanasse:2015fph}.

At LO photons couple magnetically to the nucleon magnetic moments through the Lagrangian
\begin{equation}
    \mathcal{L}_{1,0}^{mag}=\Nh^\dagger(\kappa_0+\tau_3\kappa_1)\vectS{\sigma}\cdot\Nh\vect{B},
\end{equation}
where $\kappa_0=0.43990$ ($\kappa_1=2.35295$) is the isoscalar (isovector) nucleon magnetic moment.  In this work we will use the proton and neutron magnetic moment given by the linear combinations
\begin{equation}
    \kappa_p=\kappa_0+\kappa_1\quad,\quad\kappa_n=\kappa_0-\kappa_1,
\end{equation}
respectively.  Two-nucleon magnetic currents given by the LECs $L_1$ and $L_2$ occur at NLO given by the Lagrangian
\begin{equation}
    \label{eq:L1L2def}
    \mathcal{L}_{2,1}^{mag}=\left(\frac{e}{2}L_1\hat{t}_j^\dagger\hat{s}_3\mathbf{B}_j+\mathrm{H.c.}\right)-\frac{e}{2}L_2i\epsilon^{ijk}\hat{t}_i^\dagger\hat{t}_j\mathbf{B}_k.
\end{equation}
Fitting $L_2$ to the deuteron magnetic moment we find $L_2=-1.36$~fm~\cite{Vanasse:2014sva}.  For $L_1$ we will consider several different fits in Sec.~\ref{sec:results}~\cite{Vanasse:2017kgh}. 

At LO the nucleon weak current is given by the axial vector (vector) contributions
\begin{equation}
    \label{eq:LOGTF}
    \mathcal{L}_{1,0}^{W}=-\frac{g_A}{\sqrt{2}}\Nh^\dagger\sigma_i\tau_-\Nh\Ah_i^++\frac{g_V}{\sqrt{2}}\Nh^{\dagger}\tau_-\Nh\hat{V}_0^++\mathrm{H.c.},
\end{equation}
where $g_A=1.26$ ($g_V=1$). The leptonic axial current is $\Ah_i^+$ while $V_0^{+}$ is a component of the weak leptonic vector current. We define $\tau_{\pm}=\mp\frac{1}{\sqrt{2}}(\tau_x\pm i\tau_y)$. Finally, at NLO the axial-vector (vector) receives a two-nucleon current term given by the LEC $l_{1,A}$ ($l_{1,V}$)
\begin{equation}
    \label{eq:L1AL1Vdef}
    \mathcal{L}_{2,1}^{W}=-\sqrt{2}l_{1,A}\hat{t}_k^\dagger\hat{s}_{\mp}\Ah_k^{\mp}+\sqrt{2}l_{1,V}\hat{s}_3^\dagger\hat{s}_{\mp}\hat{V}_0^{\mp}+\mathrm{H.c.}
\end{equation}
Note, $l_{1,V}$ is only necessary in the dibaryon formalism and its value, $l_{1,V}=g_Vc_{0s}$, is completely determined by the LO weak vector current contributions.\footnote{The Lagrangian for the one-nucleon F interaction [two-nucleon GT and F interactions], Eq.~\eqref{eq:LOGTF} [Eq.~\eqref{eq:L1AL1Vdef}] differs from the Lagrangian in Ref.~\cite{Nguyen:2024rlr} by a factor of $1/\sqrt{2}$ \,[$\sqrt{2}$], and by a minus sign for all GT interactions.  With these corrections to the Lagrangian the same conclusions and results of Ref.~\cite{Nguyen:2024rlr} are found.}

\section{\label{sec:2N}Two Nucleons}

\begin{figure}[hbt]
    \centering
    \includegraphics[width=0.5\linewidth]{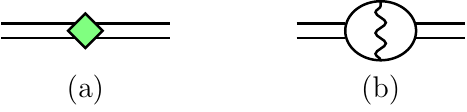}
    \caption{Diagram (b) shows the $\mathcal{O}(\alpha)$ correction to the $pp$ propagator and diagram (a) its associated counterterm.  Time flows from left to right in all diagram in this work.}
    \label{fig:ppCoulombcorrections}
\end{figure}

In \EFT the ${}^3S_1$ and ${}^1S_0$ dibaryon propagators are given by an infinite sum of diagrams yielding 
\begin{align}
    \bar{D}_{\{s,t\}}(p_0,p)=&\frac{1}{\gamma_{\{t,s\}}-\sqrt{\frac{1}{4}p^2-M_Np_0-i\epsilon}}\\\nonumber
    &\times\left[\underbrace{1\vphantom{(Z_{\{t,s\}}-1)\left(\gamma_{\{t,s\}}+\sqrt{\frac{1}{4}p^2-M_Np_0-i\epsilon}\right)}}_{\mathrm{LO}}+\underbrace{(Z_{\{t,s\}}-1)\left(\gamma_{\{t,s\}}+\sqrt{\frac{1}{4}p^2-M_Np_0-i\epsilon}\,\,\right)}_{\mathrm{NLO}}\right].
\end{align}
Following Ref.~\cite{Konig:2015aka} we note that for sufficiently large momentum $\alpha M_N\lesssim|1/a_C|\ll Q\ll \Lambda_{\not{\pi}}$, where $\Lambda_{\not{\pi}}\sim m_\pi$ is the \EFT breakdown scale and $a_C$ is the Coulomb modified $pp$ scattering length, that Coulomb corrections can be included perturbatively.  However, unlike Ref.~\cite{Konig:2015aka} where they started the $^1S_0$ channel in the unitary limit for their perturbative expansion we start with the scattering length in the physical limit such that our LO dibaryon propagator for the $^1S_0$ $pp$ channel is simply
\begin{equation}
    \bar{D}_{pp}(p_0,p)=\bar{D}_s(p_0,p).
\end{equation}
The $\mathcal{O}(\alpha)$ Coulomb corrections to the $pp$ propagator are given by the sum of diagrams in Fig.~\ref{fig:ppCoulombcorrections}~\cite{Konig:2015aka}
where diagram (a) is a counterterm to absorb the logarithmic divergence from diagram (b).  Including these diagrams the sum  of the LO and $\mathcal{O}(\alpha)$ Coulomb correction to the $pp$ propagator is given by
\begin{align}
    &i\bar{D}_{pp}(p_0,p)=i\bar{D}_s(p_0,p)\\\nonumber
    &+i\bar{D}_s(p_0,p)\left[iC_{pp}-i\alpha M_N\left(\ln\left(\frac{\alpha M_N}{2\sqrt{\frac{1}{4}p^2-M_Np_0-i\epsilon}}\right)+\ln\left(\frac{\Lambda}{\alpha M_N}\right)\right)\right]i\bar{D}_s(p_0,p),
\end{align}
where $C_{pp}$ is the counterterm from diagram (a) in Fig.~\ref{fig:ppCoulombcorrections} and $\Lambda$ comes from using a finite cutoff in the loop integral for diagram-(b).  All terms that go to zero in the limit $\Lambda\to\infty$ are dropped.  The $pp$ scattering amplitude is related to the $pp$ propagator by~\cite{Kong:1999sf}\footnote{Note we use $\hat{S}=1+i\hat{T}$ instead of of $\hat{S}=1-i\hat{T}$ as in Ref.~\cite{Konig:2015aka} for the relationship between the scattering matrix $\hat{S}$ and the $\hat{T}$ matrix.}
\begin{align}
    \label{eq:pptoTsc}
    iT_{SC}(p)=-iy^2C_\eta^2e^{i2\sigma_0}\bar{D}_{pp}\left(-\frac{p^2}{M_N},0\right),
\end{align}
where the subscript $SC$ denotes strong and Coulomb interactions are mixed.  $C_{\eta}$ is the Sommerfeld parameter defined by
\begin{equation}
    \label{eq:Ceta}
    C_\eta^2=\frac{2\pi\eta}{e^{2\pi\eta}-1},
\end{equation}
where $\eta=\alpha M_N/(2p)$.  $\sigma_0$ is the pure Coulomb $S$-wave phase shift given by
\begin{equation}
    e^{2i\sigma_0}=\frac{\Gamma(1+i\eta)}{\Gamma(1-i\eta)}.
\end{equation}
Using Eq.~\eqref{eq:pptoTsc} and resumming the $\mathcal{O}(\alpha)$ Coulomb correction to the $pp$ propagator to all orders the $pp$ scattering amplitude is given by
\begin{align}
    \label{eq:Tscpp}
    T_{SC}(p)=-\frac{4\pi}{M_N}C_\eta^2e^{2i\sigma_0}\frac{1}{\gamma_s+ip+C_{pp}-\alpha M_N\ln(i\eta)-\alpha M_N\ln\left(\frac{\Lambda}{\alpha M_N}\right)}.
\end{align}
From Ref.~\cite{Konig:2015aka} the $pp$ scattering amplitude expanded to $\mathcal{O}(\alpha)$ in the effective range expansion~\cite{Bethe:1949yr} is given by
\begin{equation}
    \label{eq:TscppERE}
    T_{SC}(p)=\frac{4\pi}{M_N}\frac{C_{\eta}^2 e^{2i\sigma_0}}{-\frac{1}{a_C}+\frac{1}{2}r_Cp^2-ip+\alpha M_N\gamma_E+\alpha M_N\ln(i\eta)},
\end{equation}
where the Coulomb modified scattering length (effective range) is found to be $a_C=-7.81$~fm ($r_C=2.79$~fm)~\cite{Bergervoet:1988zz} and $\gamma_E=0.57721\ldots$ is the Euler-Mascheroni constant.  Matching Eqs.~\eqref{eq:Tscpp} and \eqref{eq:TscppERE} the counterterm $C_{pp}$ is
\begin{equation}
    C_{pp}=\frac{1}{a_C}-\gamma_s-\alpha M_N\gamma_E+\alpha M_N\ln\left(\frac{\Lambda}{\alpha M_N}\right),
\end{equation}
and the resulting $pp$ propagator to $\mathcal{O}(\alpha)$ is
\begin{align}
    &i\bar{D}_{pp}(p_0,p)=i\bar{D}_s(p_0,p)\\\nonumber
    &+i\bar{D}_s(p_0,p)i\left[\frac{1}{a_C}-\gamma_s-\alpha M_N\gamma_E-\alpha M_N\ln\left(\frac{\alpha M_N}{2\sqrt{\frac{1}{4}p^2-M_Np_0-i\epsilon}}\right)\right]i\bar{D}_s(p_0,p).
\end{align}

\section{\label{sec:3N}Three Nucleons}

\subsection{LO and NLO}

To determine properties of three-nucleon systems, the wavefunction or equivalently the three-nucleon vertex function is required.  The three-nucleon vertex function at LO and NLO is given by the integral equation
\begin{align}
\Gb_n(p)=\oneb\delta_{n0}+\sum_{m=1}^{n}\mathbf{R}_{m}(E_B,p)\Gb_{n-m}(p)+\mathbf{K}_0(E_B,p,q)\otimes_q\Gb_{n}(q),
\label{eq:G}
\end{align}
where $n=0$ ($n=1$) is the LO (NLO correction to the) vertex function.  $\Gb_{n}(q)$ is a three dimensional cluster configuration (c.c.) space~\cite{Griesshammer:2004pe} vector defined by
\begin{equation}
    \Gb_{n}(p)=\left(\begin{array}{c}
    \G_{n,Nd}(p)\\
    \G_{n,Ns_0}(p)\\
    \G_{n,Ns_\pm}(p)
    \end{array}\right),
\end{equation}
where $\G_{n,Nd}(p)$ corresponds to an outgoing deuteron and nucleon, $\G_{n,Ns_0}(p)$ to an outgoing nucleon and neutron-proton spin singlet dibaryon, and $\G_{n,Ns_\pm}(p)$ to an outgoing nucleon and $pp$ or $nn$ dibaryon for $\jjvHe$ and $\jjvH$ respectively.  In Eq.~\eqref{eq:G} the vertex function is evaluated in the center-of-mass (c.m.) frame between dibaryon and nucleon, with $\vec{\boldsymbol{p}}$ being the outgoing deuteron momentum.  $\mathbf{K}_0(E_B,p,q)$, the kernel, is a matrix in (c.c.) space defined by 
\begin{equation}
\mathbf{K}_0(E_B,p,q)=\mathbf{R}_0(E_B,p,q)\Db(E_B,q),
\end{equation}
where
\begin{equation}
    \mathbf{R}_0(E_B,p,q)=-\frac{2\pi}{qp}Q_0\left(\frac{q^2+p^2-M_NE_B}{qp}\right)\left(\begin{array}{rrr}
    1 & -3 & -3\\
    -1 & -1 & 1\\
    -2 & 2 & 0
    \end{array}\right).
\end{equation}
$Q_0(a)$ is a Legendre function of the second kind defined by
\begin{equation}
    Q_0(a)=\frac{1}{2}\ln\left(\frac{1+a}{a-1}\right).
\end{equation}
$\Db(E_B,q)$ is the dibaryon matrix in c.c.~space given by
\begin{equation}
    \Db(E_B,q)=\left(\begin{array}{ccc}
    D_t(E_B,q) & 0 & 0 \\
    0 & D_s(E_B,q) & 0\\
    0 & 0 & D_s(E_B,q)
    \end{array}\right)=\left(\begin{array}{ccc}
        \frac{1}{\gamma_t-d(q)} & 0 & 0  \\
         0 & \frac{1}{\gamma_s-d(q)} & 0\\
         0 & 0 & \frac{1}{\gamma_s-d(q)}
    \end{array}\right),
\end{equation}
where
\begin{equation}
    d(q)=\sqrt{\frac{3}{4}q^2-M_NE_B}.
\end{equation}
$\oneb$ is a vector in c.c.~space defined as
\begin{equation}
    \oneb=\left(\begin{array}{r}
    1 \\
    -\frac{1}{3}\\
    -\frac{2}{3}
    \end{array}\right),
\end{equation}
and $\mathbf{R}_1(E_B,p)$, containing range corrections, is a matrix in c.c.~space given by
\begin{equation}
    \mathbf{R}_1(E_B,p)=\left(\begin{array}{ccc}
    \frac{c_{0t}^{(0)}}{M_N}\left(\gamma_t+d(p)\,\right) & 0 & 0\\
    0 & \frac{c_{0s}^{(0)}}{M_N}\left(\gamma_s+d(p)\,\right) &0\\
    0 & 0 & \frac{c_{0s}^{(0)}}{M_N}\left(\gamma_s+d(p)\,\right)
    \end{array}\right).
\end{equation}
In this work it is necessary to use a 3$\times$3 c.c.~space as Coulomb corrections will explicitly break isospin symmetry. The ``$\otimes$" notation is a shorthand for integration defined by
\begin{equation}
    A(q)\otimes_qB(q)=\frac{1}{2\pi^2}\int_0^\Lambda\!\! dqq^2A(q)B(q).
\end{equation}

\subsection{Coulomb Corrections}

\begin{figure}[hbt]
    \centering
    \includegraphics[width=0.75\linewidth]{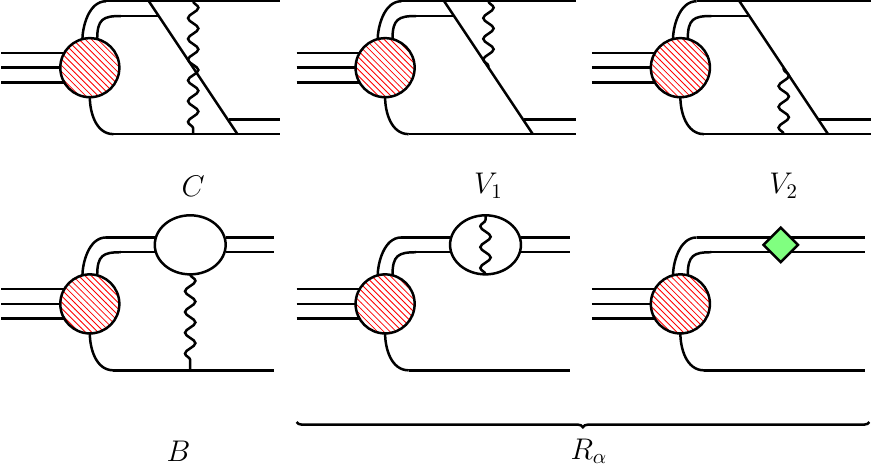}
    \caption{Coulomb correction diagrams to the inhomogeneous term of the Coulomb corrected three-nucleon vertex function.  The single line represents a nucleon, the double line a dibaryon propagator, the triple line a three-nucleon propagator, the wavy line a Coulomb photon, and the hatched circle a LO three-nucleon vertex function. All other notation is the same as in Fig.~\ref{fig:ppCoulombcorrections}.}
    \label{fig:Coulomb-inhom}
\end{figure}

The $\mathcal{O}(\alpha)$ Coulomb correction to the LO vertex function is given by the integral equation
\begin{align}
    \label{eq:galpha}
    &\Gb_{\alpha}(p)=\mathbf{R}_{\alpha}(E_B,p)\Gb_0(p)+\mathbf{K}_{\alpha}(E_B,p,q)\otimes_q\Gb_{0}(q)+\mathbf{K}_0(E_B,p,q)\otimes_q\Gb_{\alpha}(q),
\end{align}
where the inhomogeneous term is given by the sum of diagrams in Fig.~\ref{fig:Coulomb-inhom}. 
 Coulomb diagrams in Fig.~\ref{fig:Coulomb-inhom}, with the exception of the $R_\alpha$ diagrams, have been calculated previously in Ref.~\cite{Vanasse:2014kxa} for $pd$ scattering, in which the vertex function is replaced with the scattering amplitude.  The kernel $\mathbf{K}_{\alpha}(E_B,p,q)$ is given by the sum of two parts
\begin{equation}
    \mathbf{K}_{\alpha}(E_B,p,q)=\mathbf{K}_{\alpha}^{(SC)}(E_B,p,q)+\mathbf{K}_{\alpha}^{(C)}(E_B,p,q),
\end{equation}
where $\mathbf{K}_{\alpha}^{(SC)}(E_B,p,q)$ is from diagrams that mix strong and Coulomb interactions between dibaryon and nucleons and is given by the sum of the first three diagrams in Fig.~\ref{fig:Coulomb-inhom}, which yields
\begin{equation}
    \mathbf{K}_{\alpha}^{(SC)}(E_B,p,q)=\left(\begin{array}{ccc}
    C(E_B,p,q) & -3C(E_B,p,q) & -3V_1(E_B,p,q)\\
    -C(E_B,p,q) & -C(E_B,p,q) & V_1(E_B,p,q)\\ 
    -2V_2(E_B,p,q) & 2V_2(E_B,p,q) & 0
    \end{array}\right)\Db(E_B,q).
\end{equation}
$C(E_B,p,q)$, coming from the $(C)$ diagram in Fig.~\ref{fig:Coulomb-inhom}, is given by
\begin{equation}
    C(E_B,p,q)=-\frac{4\pi\alpha M_N}{qp}F_1\left[i\sqrt{3q^2+3p^2-2M_NE_B},2(d(q)+d(p)),q,p\right],
\end{equation}
where for $\mathrm{Re}[b]>\mathrm{Re}[a]$~\cite{Vanasse:2014kxa}
\begin{align}
    &F_1[a,b,c,d]=\frac{1}{4a}\left\{\ln\left(\frac{b+a}{b-a}\right)\ln(z^2+a^2)\right.\\\nonumber
    &\hspace{2cm}\left.-2\mathrm{Re}\left[\mathrm{Li}_2\left(-i\frac{z+ia}{a+b}\right)\right]+2\mathrm{Re}\left[\mathrm{Li}_2\left(i\frac{z+ia}{b-a}\right)\right]\right\}\Big{|}^{c+d}_{|c-d|},
\end{align}
and for $\mathrm{Re}[a]>\mathrm{Re}[b]$
\begin{align}
    &F_1[a,b,c,d]=\frac{1}{a}\tan^{-1}\left(\frac{z}{a}\right)\tan^{-1}\left(\frac{z}{b}\right)\Big{|}^{c+d}_{|c-d|}-\frac{1}{4a}\left\{\ln\left(\frac{b+a}{a-b}\right)\ln(z^2+b^2)\right.\\\nonumber
    &\left.-2\mathrm{Re}\left[\mathrm{Li}_2\left(-i\frac{z+ib}{a+b}\right)\right]+2\mathrm{Re}\left[\mathrm{Li}_2\left(i\frac{z+ib}{a-b}\right)\right]\right\}\Big{|}^{c+d}_{|c-d|}.
\end{align}
The bar notation is defined by
\begin{equation}
    f(z)\Big{|}^{c+d}_{|c-d|}=f(c+d)-f(|c-d|).
\end{equation}
$V_1(E_B,p,q)$, coming from the $(V_1)$ diagram in Fig.~\ref{fig:Coulomb-inhom}, is given by
\begin{equation}
    V_1(E_B,p,q)=\frac{4\pi\alpha M_N}{qp}F_1\left[2d(q),2d(q),q,2p\right],
\end{equation}
and by time reversal symmetry $V_2(E_B,p,q)$, coming from the $(V_2)$ diagram in Fig.~\ref{fig:Coulomb-inhom}, by
\begin{equation}
    V_2(E_B,p,q)=V_1(E_B,q,p).
\end{equation}
For $V_1(E_B,p,q)$ the function $F_1[a,b,c,d]$ has the first two arguments equal, $a=b$.  In this limit the function $F_1[a,b,c,d]$ simplifies to
\begin{equation}
    F_1[a,a,c,d]=\frac{1}{2a}\left[\left(\tan^{-1}\left(\frac{c+d}{a}\right)\right)^2-\left(\tan^{-1}\left(\frac{|c-d|}{a}\right)\right)^2\right].
\end{equation}
The kernel $\mathbf{K}_{\alpha}^{(C)}(E_B,p,q)$ is defined by
\begin{equation}
    \mathbf{K}_{\alpha}^{(C)}(E_B,p,q)=\left(\begin{array}{ccc}
    B(E_B,p,q) & 0 & 0\\
    0 & B(E_B,p,q)& \\ 
    0 & 0 & 0
    \end{array}\right)\Db(E_B,q).
\end{equation}
$B(E_B,p,q)$, coming from the $(B)$ diagram in Fig.~\ref{fig:Coulomb-inhom}, is given by\footnote{$C(E_B,p,q)$, $V_1(E_B,p,q)$, $V_2(E_B,p,q)$, and $B(E_B,p,q)$ differ from Ref.~\cite{Vanasse:2014kxa} by a factor of $\pi$.}
\begin{equation}
    B(E_B,p,q)=\frac{4\pi\alpha M_N}{qp}F_1\left[0,2(d(q)+d(p)),q,p\right].
\end{equation}
The form of $F_1[a,b,c,d]$ in the case $a=0$ simplifies to
\begin{equation}
    F_1[0,b,q,p]=\frac{1}{b}Q_0\left(\frac{q^2+p^2}{2qp}\right)-\frac{1}{b}Q_0\left(\frac{q^2+p^2+b^2}{2qp}\right)-\frac{1}{z}\tan^{-1}\left(\frac{z}{b}\right)\Big{|}^{c+d}_{|c-d|}.
\end{equation}
This function has a logarithmic divergence when $q=p$.  These divergences are integrable and to deal with them we use a subtraction technique to lessen their effect.  Typically, these divergences are regulated by giving the photon a finite mass.  However, given that this divergence only occurs in the inhomogeneous term of our integral equation for $\Gb_{\alpha}(q)$ we can instead use a subtraction technique to address this divergence.  Finally, the function $\mathbf{R}_{\alpha}(E_B,p)$ in Eq.~\eqref{eq:galpha}, coming from the $R_\alpha$ diagrams in Fig.~\ref{fig:Coulomb-inhom}, is given by
\begin{equation}
    \mathbf{R}_{\alpha}(E_B,p)=\left[-\frac{1}{a_C}+\gamma_s+\alpha M_N\gamma_E+\alpha M_N\ln\left(\frac{\alpha M_N}{2d(p)}\right)\right]\left(\begin{array}{ccc}
    0 & 0 & 0\\[-2mm]
    0 & 0 & 0\\[-2mm]
    0 & 0 & 1
    \end{array}\right)\Db(E_B,p).
\end{equation}

In addition to the vertex function we also need the three-nucleon wavefunction renormalization, which is given by the residue about the triton pole of the triton propagator, yielding (See Ref.~\cite{Vanasse:2015fph} for details)
\begin{equation}
    \sqrt{Z_\psi}=\sqrt{\frac{\pi}{\Sigma_0'(E_B)}}\left[\underbrace{1\vphantom{\frac{1}{2}\frac{\Sigma_1'(E_B)}{\Sigma_0'(E_B)}}}_{\mathrm{LO}}-\underbrace{\frac{1}{2}\frac{\Sigma_1'(E_B)}{\Sigma_0'(E_B)}}_{\mathrm{NLO}}-\underbrace{\frac{1}{2}\frac{\Sigma_\alpha'(E_B)}{\Sigma_0'(E_B)}}_{\mathcal{O}(\alpha)}+\cdots\right]
\end{equation}
where 
\begin{equation}
    \Sigma_n(E)=-\pi\bar{\boldsymbol{1}}^T\Db(E,q)\otimes_q\Gb_n(E,q),
\end{equation}
and
\begin{equation}
    \Sigma_\alpha(E)=-\pi\bar{\boldsymbol{1}}^T\Db(E,q)\otimes_q\Gb_\alpha(E,q).
\end{equation}
The c.c.~space vector $\bar{\boldsymbol{1}}$ is 
\begin{equation}
    \bar{\boldsymbol{1}}=\left(\begin{array}{r}
    1 \\ 
    -1\\
    -1
    \end{array}\right).
\end{equation}
With the wavefunction renormalization known, the renormalized LO three-nucleon vertex function is
\begin{equation}
    \widetilde{\Gammab}_0(p)=\sqrt{\frac{\pi}{\Sigma_0'(E_B)}}\Gb_0(E_B,p),
\end{equation}
the renormalized NLO correction to the three-nucleon vertex function is
\begin{equation}
    \widetilde{\Gammab}_1(p)=\sqrt{\frac{\pi}{\Sigma_0'(E_B)}}\left[\Gb_1(E_B,p)-\frac{1}{2}\frac{\Sigma_1'(E_B)}{\Sigma_0'(E_B)}\Gb_0(E_B,p)\right],
\end{equation}
and the renormalized $\mathcal{O}(\alpha)$ Coulomb correction to the vertex function is
\begin{equation}
    \widetilde{\Gammab}_\alpha(p)=\sqrt{\frac{\pi}{\Sigma_0'(E_B)}}\left[\Gb_\alpha(E_B,p)-\frac{1}{2}\frac{\Sigma_\alpha'(E_B)}{\Sigma_0'(E_B)}\Gb_0(E_B,p)\right].
\end{equation}
The $\mathcal{O}(\alpha)$ correction to the $\jjvHe$ binding energy is~\cite{Vanasse:2015fph}
\begin{equation}
    B_{1}^{(\alpha)}=-\frac{\Sigma_\alpha(E_B)}{\Sigma_0'(E_B)}.
\end{equation}
For convenience in calculations of the form factors we also define
\begin{equation}
    \label{eq:gammadef}
    \Gammab_{n}(p)=\Db(E_B,p)\widetilde{\Gammab}_{n}(p)\quad,\quad\Gammab_{\alpha}(p)=\Db(E_B,p)\widetilde{\Gammab}_{\alpha}(p).
\end{equation}

\section{\label{sec:form}Form Factors}

\subsection{LO}

\begin{figure}[hbt]
    \centering
    \includegraphics[width=0.75\linewidth]{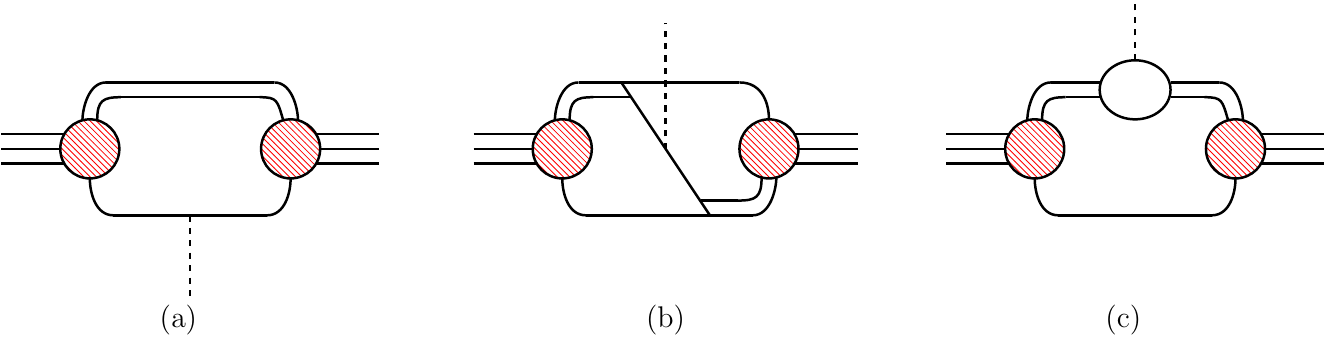}
    \caption{Diagrams for the generic LO form factor in \EFT.  The dashed line represents the external current and all other notation is the same as in Fig.~\ref{fig:Coulomb-inhom}}
    \label{fig:LOFF}
\end{figure}

The LO generic three-nucleon form factor (valid for any current with non-derivative coupling) is given by the sum of diagrams in Fig.~\ref{fig:LOFF}~\cite{Vanasse:2017kgh}.  
Evaluation of the generic three-nucleon form factor is done in the Breit frame, in which the current imparts momentum but no energy to the nuclear system.  The three-nucleon system has total momentum $\vect{K}$ ($\vect{P}$) before (after) interacting with the current.  $\vect{Q}=\vect{P}-\vect{K}$ is the total momentum of the current and form factors are functions of $Q^2$. The total energy of the three-nucleon system before and after the current is $E=E_B+\frac{1}{6M_N}K^2$.  Summing the diagrams in Fig.~\ref{fig:LOFF} yields 
\begin{align}
    \label{eq:LOForm}
    &Z_{\psi}^{\mathrm{LO}}\sum_{j=a,b,c}\int\!\!\frac{d^4k}{(2\pi)^4}\int\!\!\frac{d^4p}{(2\pi)^4}\Gb_0(E,\vect{P},p_0,\vect{p})\boldsymbol{\chi}_{j}(E,\vect{K},\vect{P},p_0,k_0,\vect{p},\vect{k})\Gb_0(E,\vect{K},k_0,\vect{k}),
\end{align}
where $\Gb_0(E,\vect{K},k_0,\vect{k})$ is the boosted LO vertex function given by
\begin{align}
    \Gb_0(E,\vect{K},k_0,\vect{k})=\oneb+\left[\mathbf{R}_0\left(\frac{2}{3}E_B+k_0-\frac{\vect{K}\cdot\vect{k}}{3M_N}+\frac{k^2}{2M_N},k,q\right)\Db(E_B,q)\right]\otimes_q\Gb_0(q).
\end{align}
The functions $\boldsymbol{\chi}_{j}(E,\vect{K},\vect{P},p_0,k_0,\vect{p},\vect{k})$ have been given previously in 2$\times$2 c.c.~space for the triton charge form factor~\cite{Vanasse:2015fph}.  Slightly modified forms of $\boldsymbol{\chi}_{j}(E,\vect{K},\vect{P},p_0,k_0,\vect{p},\vect{k})$ in 3$\times$3 c.c.~space for our form factors of interest are given in App.~\ref{app:chi}.

Although in principle the form factor can be calculated as a function of $Q^2$ we take the additional simplifying assumption $Q^2=0$ and focus on the zeroth order moment of the form factor in this work.  The LO generic three-nucleon form factor at $Q^2=0$ is given by~\cite{Vanasse:2017kgh}
\begin{align}
    \label{eq:LOF}
    F_0(0)=2\pi M_N\Gammab_0(p)\otimes_p\left\{\frac{\pi\delta(p-k)}{2k^2d(k)}\Mb_1-\frac{1}{(p^2+k^2-M_NE_B)^2-p^2k^2}\Mb_2\right\}\otimes_k\Gammab_0(k).
\end{align}
Using isospin symmetry the forms of the matrices $\Mb_1$ and $\Mb_2$ have been calculated previously in the 2$\times$2 c.c.~space for charge, magnetic~\cite{Vanasse:2017kgh}, and GT and F form factors~\cite{Nguyen:2024rlr}.  However, given that Coulomb interactions break isospin symmetry it is necessary to calculate 3$\times$3 c.c.~space matrices  in which different spin singlet states are distinguished.  The 3$\times$3 c.c.~space $\Mb_1$ matrix elements for the magnetic and charge $\jjvHe$ form factors as well as the GT and F form factors are given in Table~\ref{tab:M1LO}.
\begin{table}
    \begin{tabular}{|c|c|c|c|c|c|c|c|c|c|}
        \hline
       &$\left[\Mb_1\right]_{11}$   & $\left[\Mb_1\right]_{12}$ & $\left[\Mb_1\right]_{13}$ & $\left[\Mb_1\right]_{21}$ & $\left[\Mb_1\right]_{22}$ & $\left[\Mb_1\right]_{23}$ & $\left[\Mb_1\right]_{31}$ & $\left[\Mb_1\right]_{32}$ & $\left[\Mb_1\right]_{33}$ \\\hline
       $F_C^{\jjvHe}(0)$ & 2 & 0 & 0 & 0 & 6 & 0 & 0 & 0 & 3\\\hline
       $F_M^{\jjvHe}(0)$ & $\frac{\kappa_p+2\kappa_n}{3}$ & $\kappa_p-\kappa_n$ & 0 & $\kappa_p-\kappa_n$ & $3\kappa_p$ & 0 & 0 & 0 & $\frac{3}{2}\kappa_n$\\\hline
       $F_W^{GT}(0)$ & $\frac{1}{3}$ & 0 & -1 & 0 & 3 & 0 & -1 & 0 & 0\\\hline
       $F_W^{F}(0)$ & 1 & 0 & 0 & 0 & -3 & 3 & 0 & 3 & 0\\\hline
    \end{tabular}
    \caption{\label{tab:M1LO} Matrix elements of $\Mb_1$ for the $\jjvHe$ charge and magnetic form factors and the GT and F form factors.}
\end{table}

Table~\ref{tab:M2LO} gives the same information as Table~\ref{tab:M1LO} but for the 3$\times$3 c.c.~space $\Mb_2$ matrix elements.
\begin{table}
    \begin{tabular}{|c|c|c|c|c|c|c|c|c|c|}
        \hline
       &$\left[\Mb_2\right]_{11}$   & $\left[\Mb_2\right]_{12}$ & $\left[\Mb_2\right]_{13}$ & $\left[\Mb_2\right]_{21}$ & $\left[\Mb_2\right]_{22}$ & $\left[\Mb_2\right]_{23}$ & $\left[\Mb_2\right]_{31}$ & $\left[\Mb_2\right]_{32}$ & $\left[\Mb_2\right]_{33}$ \\\hline
       $F_C^{\jjvHe}(0)$ & -2 & 6 & 6 & 6 & 6 & -6 & 6 & -6 & 0\\\hline
       $F_M^{\jjvHe}(0)$ & $\frac{2\kappa_p-5\kappa_n}{3}$ & $2\kappa_p+\kappa_n$ & $3\kappa_n$ & $2\kappa_p+\kappa_n$ & $6\kappa_p-3\kappa_n$ & $-3\kappa_n$ & $3\kappa_n$ & $-3\kappa_n$ & 0\\\hline
       $F_W^{GT}(0)$ & -$\frac{7}{3}$ & 3 & 1 & 3 & 3 & -3 & 1 & -3 & -3\\\hline
       $F_W^{F}(0)$ & -1 & 3 & 3 & 3 & -9 & 3 & 3 & 3 & -3\\\hline
    \end{tabular}
    \caption{\label{tab:M2LO} Matrix elements of $\Mb_2$ for the $\jjvHe$ charge and magnetic form factors and the GT and F form factors.}
\end{table}

\subsection{NLO corrections}

\begin{figure}[hbt]
    \centering
    \includegraphics[width=0.75\linewidth]{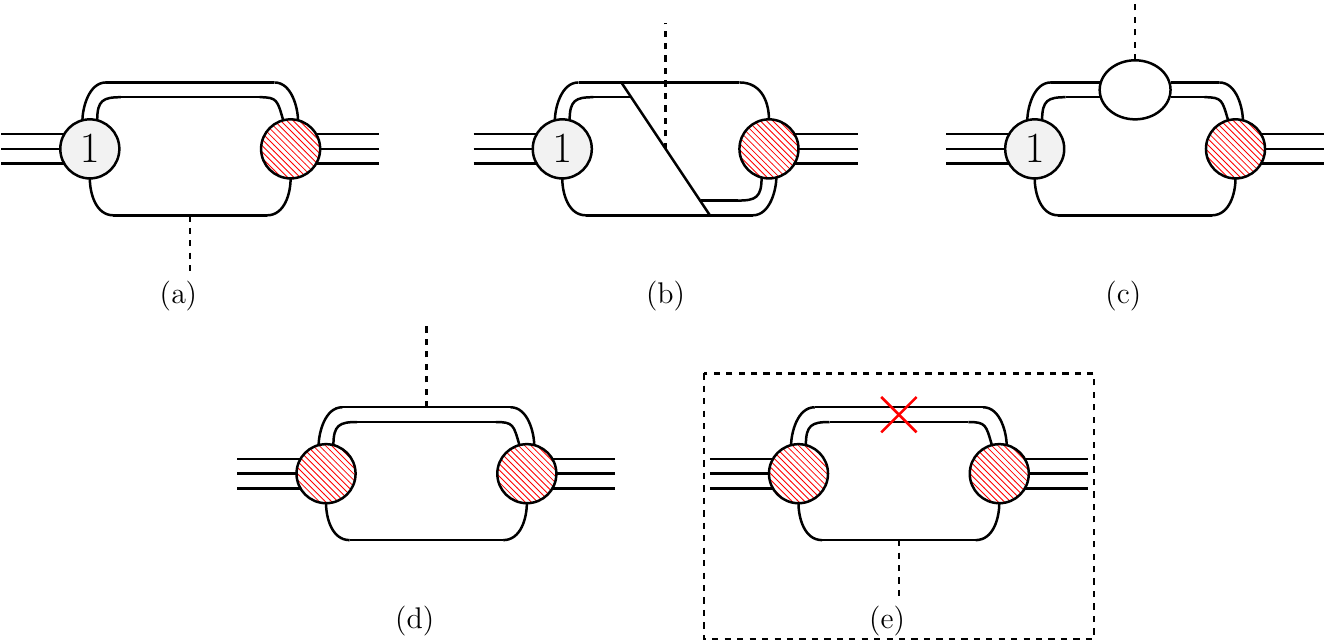}
    \caption{The NLO correction to the generic three-nucleon form factor. The circle with a ``1" represents the NLO correction to the three-nucleon vertex function and the red \textcolor{red}{$\times$} is a range correction.  Diagram (e) is boxed because it is subtracted to avoid double counting.  All other notation is the same as in Figs.~\ref{fig:Coulomb-inhom} and ~\ref{fig:LOFF}.  Diagrams related by time reversal symmetry are not shown.}
    \label{fig:NLOFF}
\end{figure}

The NLO correction to the generic three-nucleon form factor is given by the sum of diagrams in Fig.~\ref{fig:NLOFF}.  Diagram (e) is boxed because it is subtracted to avoid double counting from diagram (a) and its time reversed version.  Diagram (d) for the charge form factor comes from gauging the dibaryon kinetic term and for the magnetic form factor from the LECs in Eq.~\eqref{eq:L1L2def}.  The GT and F form factor diagram (d) comes from the LECs in Eq.~\eqref{eq:L1AL1Vdef}.
The sum of these diagrams is given by
\begin{align}
    \label{eq:NLOForm}
    &Z_{\psi}^{\mathrm{LO}}\sum_{j=a,b,c}\int\!\!\frac{d^4k}{(2\pi)^4}\int\!\!\frac{d^4p}{(2\pi)^4}\Gb_1(E,\vect{P},p_0,\vect{p})\boldsymbol{\chi}_{j}(E,\vect{K},\vect{P},p_0,k_0,\vect{p},\vect{k})\Gb_0(E,\vect{K},k_0,\vect{k})\\\nonumber
    &+Z_{\psi}^{\mathrm{LO}}\sum_{j=a,b,c}\int\!\!\frac{d^4k}{(2\pi)^4}\int\!\!\frac{d^4p}{(2\pi)^4}\Gb_0(E,\vect{P},p_0,\vect{p})\boldsymbol{\chi}_{j}(E,\vect{K},\vect{P},p_0,k_0,\vect{p},\vect{k})\Gb_1(E,\vect{K},k_0,\vect{k})\\\nonumber
    &+Z_{\psi}^{\mathrm{LO}}\sum_{j=d,e}\int\!\!\frac{d^4k}{(2\pi)^4}\int\!\!\frac{d^4p}{(2\pi)^4}\Gb_0(E,\vect{P},p_0,\vect{p})\boldsymbol{\chi}_{j}(E,\vect{K},\vect{P},p_0,k_0,\vect{p},\vect{k})\Gb_0(E,\vect{K},k_0,\vect{k}),
\end{align}
where the boosted NLO correction to the vertex function is given by
\begin{align}
    \Gb_1(E,\vect{K},k_0,\vect{k})=&\mathbf{R}_1\left(\frac{2}{3}E+k_0+\frac{1}{2M_N}\left(\vect{k}+\frac{2}{3}\vect{K}\right)^2,\vect{k}+\frac{2}{3}\vect{K}\right)\Gb_0(E,\vect{K},k_0,\vect{k})\\\nonumber
    &+\left[\mathbf{R}_0\left(\frac{2}{3}E_B+k_0-\frac{\vect{K}\cdot\vect{k}}{3M_N}+\frac{k^2}{2M_N},k,q\right)\Db(E_B,q)\right]\otimes_q\Gb_1(q).
\end{align}
$\boldsymbol{\chi}_{j}(E,\vect{K},\vect{P},p_0,k_0,\vect{p},\vect{k})$ for $j=d$ and $j=e$ is given in App.~\ref{app:chi}.  Taking $Q^2=0$ the NLO correction to the generic form factor is
\begin{align}
    &F_1(0)=\\\nonumber
    &2\pi M_N\Gammab_0(p)\otimes_p\left\{\frac{\pi\delta(p-k)}{2k^2d(k)}\Mb_1-\frac{1}{(p^2+k^2-M_NE_B)^2-p^2k^2}\Mb_2\right\}\otimes_k\Gammab_1(k)\\\nonumber
    &+2\pi M_N\Gammab_1(p)\otimes_p\left\{\frac{\pi\delta(p-k)}{2k^2d(k)}\Mb_1-\frac{1}{(p^2+k^2-M_NE_B)^2-p^2k^2}\Mb_2\right\}\otimes_k\Gammab_0(k)\\\nonumber
   &-2\pi M_N\Gammab_0(p)\otimes_p\left\{\pi\frac{\delta(p-k)}{k^2}\Mb_3\right\}\otimes_k\Gammab_0(k).
\end{align}
The new NLO c.c.~space matrix $\Mb_3$ is defined in Tab.~\ref{tab:M3NLO} for our form factors of interest.
\begin{table}
    \begin{tabular}{|c|c|c|c|c|c|c|c|c|c|}
        \hline
       &$\left[\Mb_3\right]_{11}$   & $\left[\Mb_3\right]_{12}$ & $\left[\Mb_3\right]_{13}$ & $\left[\Mb_3\right]_{21}$ & $\left[\Mb_3\right]_{22}$ & $\left[\Mb_3\right]_{23}$ & $\left[\Mb_3\right]_{31}$ & $\left[\Mb_3\right]_{32}$ & $\left[\Mb_3\right]_{33}$ \\\hline
       &&&&&&&&&\\[-7mm]
       $F_C^{\jjvHe}(0)$ & $2\frac{c_{0t}^{(0)}}{M_N}$ & 0 & 0 & 0 & $6\frac{c_{0s}^{(0)}}{M_N}$ & 0 & 0 & 0 & $3\frac{c_{0s}^{(0)}}{M_N}$\\\hline
       &&&&&&&&&\\[-7mm]
       $F_M^{\jjvHe}(0)$ & $-\frac{1}{3}\kappa_p\frac{c_{0t}^{(0)}}{M_N}-\frac{2}{3}L_2$ & $-L_1$ & 0 & $-L_1$ & $3\kappa_p\frac{c_{0s}^{(0)}}{M_N}$ & 0 & 0 & 0 & $\frac{3}{2}\kappa_n \frac{c_{0s}^{(0)}}{M_N}$\\\hline
       &&&&&&&&&\\[-7mm]
       $F_W^{GT}(0)$ & $\frac{1}{3}\frac{c_{0t}^{(0)}}{M_N}$ & 0 & $\frac{l_{1,A}}{g_AM_N}$  & 0 & $3\frac{c_{0s}^{(0)}}{M_N}$ & 0 & $\frac{l_{1,A}}{g_AM_N}$ & 0 & 0\\\hline
       &&&&&&&&&\\[-7mm]
       $F_W^{F}(0)$ & $\frac{c_{0t}^{(0)}}{M_N}$ & 0 & 0 & 0 & $-3\frac{c_{0s}^{(0)}}{M_N}$ & $3\frac{c_{0s}^{(0)}}{M_N}$ & 0 & $3\frac{c_{0s}^{(0)}}{M_N}$ & 0\\\hline
    \end{tabular}
    \caption{\label{tab:M3NLO} Matrix elements of $\Mb_3$ for the $\jjvHe$ charge and magnetic form factors and the GT and F form factors.}
\end{table}

\subsection{Coulomb Corrections}
\begin{figure}[hbt]
    \centering
    \includegraphics[width=.75\linewidth]{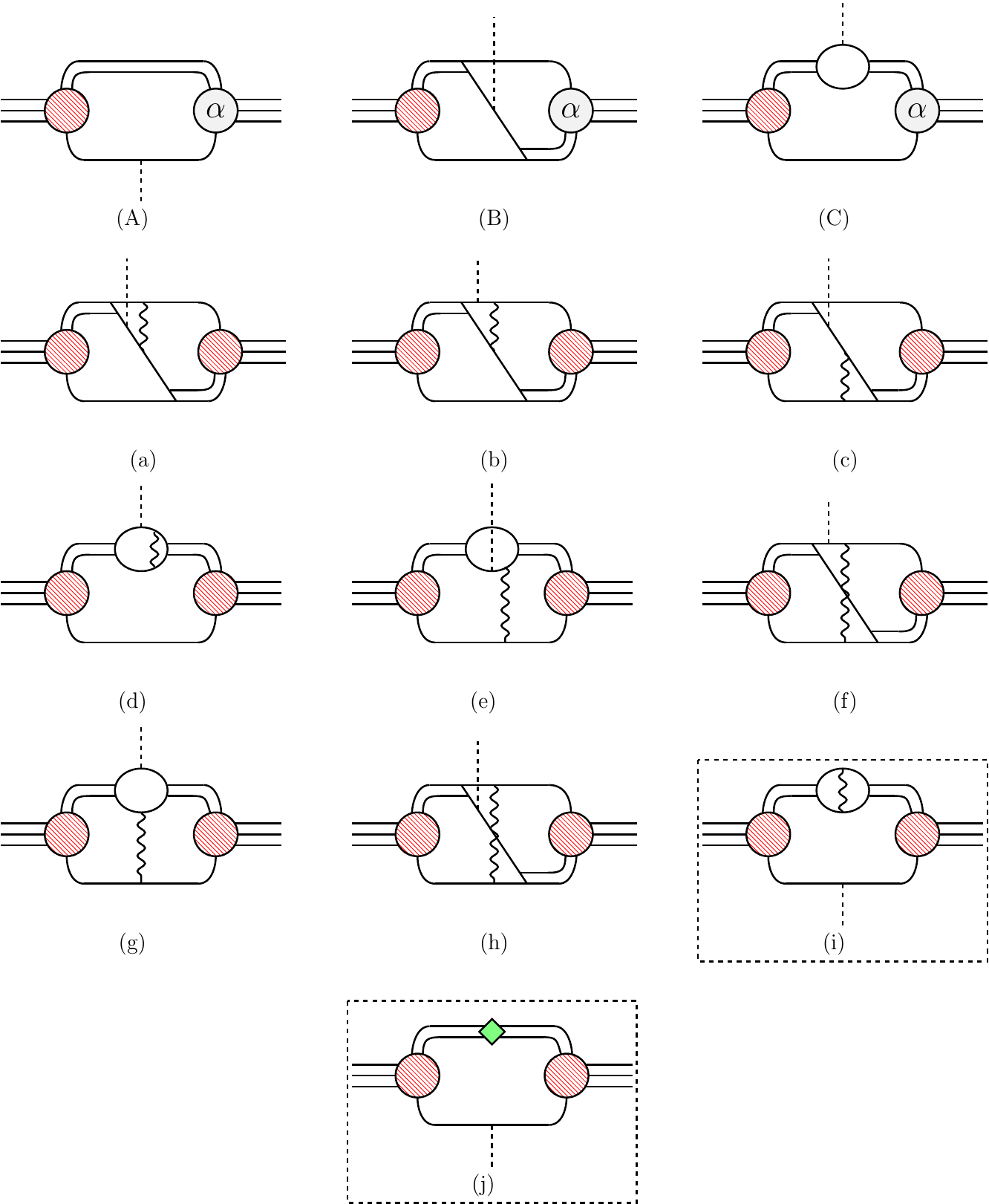}
    \caption{Diagrams for the $\mathcal{O}(\alpha)$ correction to the three-nucleon form factor.  Boxed diagrams are subtracted to avoid double counting and the circle with $\alpha$ represents the Coulomb corrected vertex function.  All other notation is the same as in Figs.~\ref{fig:ppCoulombcorrections}, \ref{fig:Coulomb-inhom}, and \ref{fig:LOFF}.}
    \label{fig:alphaFF}
\end{figure}

The $\mathcal{O}(\alpha)$ correction to the LO generic three-nucleon form factor is given by the sum of diagrams in Fig.~\ref{fig:alphaFF}, where the boxed diagrams (i) and (j) are subtracted to avoid double counting from the sum of diagram (A) and its time reversed version.
Diagrams (g) through (j) only occur for the magnetic form factor.  Also diagrams related by time reversal symmetry are not shown in Fig.~\ref{fig:alphaFF} and do not contribute to the GT or F form factor.  Summing diagrams (A), (B), and (C) in Fig.~\ref{fig:alphaFF} gives
\begin{align}
    \label{eq:alphaForm}
    &Z_{\psi}^{\mathrm{LO}}\sum_{j=a,b,c}\int\!\!\frac{d^4k}{(2\pi)^4}\int\!\!\frac{d^4p}{(2\pi)^4}\Gb_\alpha(E,\vect{P},p_0,\vect{p})\chi_{j}(E,\vect{K},\vect{P},p_0,k_0,\vect{p},\vect{k})\Gb_0(E,\vect{K},k_0,\vect{k})\\\nonumber
    &+Z_{\psi}^{\mathrm{LO}}\sum_{j=a,b,c}\int\!\!\frac{d^4k}{(2\pi)^4}\int\!\!\frac{d^4p}{(2\pi)^4}\Gb_0(E,\vect{P},p_0,\vect{p})\chi_{j}(E,\vect{K},\vect{P},p_0,k_0,\vect{p},\vect{k})\Gb_\alpha(E,\vect{K},k_0,\vect{k}),
\end{align}
where $\Gb_{\alpha}(E,\vect{K},k_0,\vect{k})$ is the boosted $\mathcal{O}(\alpha)$ Coulomb correction to the vertex function given by 
\begin{align}
    \label{eq:Galphaboost}
    \Gb_{\alpha}(E,\vect{K},k_0,\vect{k})=&\mathbf{R}_{\alpha}\left(\frac{2}{3}E+k_0+\frac{1}{2M_N}\left(\vect{k}+\frac{2}{3}\vect{K}\right)^2,\vect{k}+\frac{2}{3}\vect{K}\right)\Gb_0(E,\vect{K},k_0,\vect{k})\\\nonumber
    &+\mathbf{K}_{\alpha}(q,k,E,\vect{K},k_0)\otimes_q\Gb_{0}(q)\\\nonumber
    &+\left[\mathbf{R}_0\left(\frac{2}{3}E_B+k_0-\frac{\vect{K}\cdot\vect{k}}{3M_N}+\frac{k^2}{2M_N},k,q\right)\Db(E_B,q)\right]\otimes_q\Gb_{\alpha}(q).
\end{align}
The detailed form of $\mathbf{K}_{\alpha}(q,k,E,\vect{K},k_0)$ is lengthy and given in App.~\ref{app:boostGalpha}.  Taking the limit $Q^2=0$, the boosted $\mathcal{O}(\alpha)$ Coulomb correction to the vertex function simply becomes the $\mathcal{O}(\alpha)$ Coulomb correction to the vertex function in the c.m.~frame for diagram (B) and (C).  However, for diagram (A) a second order energy pole leads to a derivative in $k_0$ in the limit $Q^2=0$.  Thus in the limit $Q^2=0$ diagram (A) will have a contribution where the boosted $\mathcal{O}(\alpha)$ Coulomb correction to the vertex function becomes the $\mathcal{O}(\alpha)$ Coulomb correction to the vertex function in the c.m.~frame, but will also have contributions that have the same form, but different c.c.~space matrix elements, as diagrams (a) through (j). Keeping only those contributions where the boosted $\mathcal{O}(\alpha)$ Coulomb correction to the vertex function becomes the $\mathcal{O}(\alpha)$ Coulomb correction to the vertex function in the c.m.~frame, in addition to the contribution from diagrams (B) and (C), gives the contribution to the generic form factor in the limit $Q^2=0$
\begin{align}
    \label{eq:FABC}
    &F_{A+B+C}(0)=2\pi M_N\Gammab_\alpha(p)\otimes_p\left\{\frac{\pi\delta(p-k)}{2k^2d(k)}\Mb_1-\frac{1}{(p^2+k^2-M_NE_B)^2-p^2k^2}\Mb_2\right\}\otimes_k\Gammab_0(k)\\\nonumber
    &+2\pi M_N\Gammab_0(E_B,p)\otimes_p\left\{\frac{\pi\delta(p-k)}{2k^2d(k)}\Mb_1-\frac{1}{(p^2+k^2-M_NE_B)^2-p^2k^2}\Mb_2\right\}\otimes_k\Gammab_\alpha(k).
\end{align}
Matrices $\Mb_1$ and $\Mb_2$ are defined in Tabs.~\ref{tab:M1LO} and~\ref{tab:M2LO} respectively. This expression also includes parts of diagrams (i) and (j) that have been subtracted away.  The remaining parts of diagrams (i) and (j) behave like diagram (d) and are absorbed into its expression.  Both terms in Eq.~\eqref{eq:FABC} are necessary for the magnetic and charge form factor, however, given that Coulomb corrections only occur after the current for the F and GT form factor the second term in Eq.~\eqref{eq:FABC} does not exist for the F and GT form factor.  The Coulomb correction to the generic three-nucleon form factor from diagrams (a) and (b) as well as the part of diagram (A) that only differs from diagrams (a) and (b) by the c.c.~space matrix elements combine to give
\begin{align}
    &F_{a+b}(0)=\pi M_N^2\alpha\Gammab_0(p)\otimes_p\frac{1}{(p^2+k^2-M_NE_B)^2-k^2p^2}\frac{1}{d(k)}\Mb_{a+b}\otimes_k\Gammab_0(k).
\end{align}
Table~\ref{tab:Maplusb} gives the matrix elements of $\Mb_{a+b}$ for our form factors of interest.
\begin{table}[hbt]
    \begin{tabular}{|c|c|c|c|c|c|c|c|c|c|}
        \hline
       &$\left[\Mb_{a+b}\right]_{11}$   & $\left[\Mb_{a+b}\right]_{12}$ & $\left[\Mb_{a+b}\right]_{13}$ & $\left[\Mb_{a+b}\right]_{21}$ & $\left[\Mb_{a+b}\right]_{22}$ & $\left[\Mb_{a+b}\right]_{23}$ & $\left[\Mb_{a+b}\right]_{31}$ & $\left[\Mb_{a+b}\right]_{32}$ & $\left[\Mb_{a+b}\right]_{33}$ \\\hline
       Ch & 0 & 0 & 12 & 0 & 0 & -12 & 0 & 0 & 0\\\hline
       mag & 0 & 0 & $6\kappa_n$ & 0 & 0 & $-6\kappa_n$ & 0 & 0 & 0\\\hline
       GT & -2 & 0 & 0 & 2 & 0 & 0 & 0 & 0 & 0\\\hline
       F & 0 & 6 & 0 & 0 & -6 & 0 & 0 & 0 & 0\\\hline
    \end{tabular}
    \caption{\label{tab:Maplusb}Matrix elements of $\Mb_{a+b}$ for charge, magnetic, GT, and F form factors.}
\end{table}
The Coulomb correction to the generic three-nucleon form factor from diagram (c) as well as the part of diagram (A) that only differs from diagram (c) by the c.c.~space matrix elements combine to give
\begin{align}
    &F_c(0)=16\pi M_N^2\alpha\Gammab_0(p)\otimes_p\frac{1}{pk}
    F_2\left[2d(p),2d(p),p,2k\right]\Mb_c\otimes_k\Gammab_0(k).
\end{align}
$F_2[a,b,c,d]$ is a function defined by
\begin{align}
    &F_2[a,b,c,d]=\frac{1}{2a^3}\left\{\left[\tan^{-1}\left(\frac{z}{a}\right)+\frac{az}{a^2+z^2}\right]\tan^{-1}\left(\frac{z}{b}\right)\right.\\[2mm]\nonumber
    &\hspace{2cm}\left.-\frac{ab}{2(a^2-b^2)}\ln\left(\frac{z^2+b^2}{z^2+a^2}\right)\right\}\Big{|}^{c+d}_{|c-d|}-\frac{b}{2a^3}F_1[b,a,c,d],
\end{align}
and in the limit $a=b$ simplifies to
\begin{align}
    &F_2[a,a,c,d]=\frac{1}{2a^3}\left\{\left[\tan^{-1}\left(\frac{z}{a}\right)+\frac{az}{a^2+z^2}\right]\tan^{-1}\left(\frac{z}{a}\right)\right.\\[2mm]\nonumber
    &\hspace{4cm}\left.+\frac{a^2}{2(z^2+a^2)}\right\}\Big{|}^{c+d}_{|c-d|}-\frac{1}{2a^2}F_1[a,a,c,d].
\end{align}
Matrix elements for $\Mb_c$ for our form factors of interest are given in Table~\ref{tab:Mc}.
\begin{table}[hbt]
    \begin{tabular}{|c|c|c|c|c|c|c|c|c|c|}
        \hline
       &$\left[\Mb_{c}\right]_{11}$   & $\left[\Mb_{c}\right]_{12}$ & $\left[\Mb_{c}\right]_{13}$ & $\left[\Mb_{c}\right]_{21}$ & $\left[\Mb_{c}\right]_{22}$ & $\left[\Mb_{c}\right]_{23}$ & $\left[\Mb_{c}\right]_{31}$ & $\left[\Mb_{c}\right]_{32}$ & $\left[\Mb_{c}\right]_{33}$ \\\hline
       Ch & 0 & 0 & 0 & 0 & 0 & 0 & 12 & -12 & 0\\\hline
       mag & 0 & 0 & 0 & 0 & 0 & 0 & $6\kappa_n$ & -$6\kappa_n$ & 0\\\hline
       GT & 0 & 0 & 0 & 0 & 0 & 0 & 1 & -3 & -3\\\hline
       F & 0 & 0 & 0 & 0 & 0 & 0 & 3 & 3 & -3\\\hline
    \end{tabular}
    \caption{\label{tab:Mc}Matrix elements of $\Mb_{c}$ for charge, magnetic, GT, and F form factors.}
\end{table}
The Coulomb correction to the generic three-nucleon from factor from diagram (d) as well as the part of diagram (A) minus diagrams (i) and (j) that only differs from diagram (d) by the c.c.~space matrix elements combine to give
\begin{align}
    &F_d(0)=\frac{M_N^2\alpha}{8}\Gammab_0(k)\otimes_k\frac{1}{d^2(k)}\Mb_d\Gammab_0(k).
\end{align}
Matrix elements of $\Mb_d$ for our form factors of interest are given in Table~\ref{tab:Md}.
\begin{table}[hbt]
    \begin{tabular}{|c|c|c|c|c|c|c|c|c|c|}
        \hline
       &$\left[\Mb_{d}\right]_{11}$   & $\left[\Mb_{d}\right]_{12}$ & $\left[\Mb_{d}\right]_{13}$ & $\left[\Mb_{d}\right]_{21}$ & $\left[\Mb_{d}\right]_{22}$ & $\left[\Mb_{d}\right]_{23}$ & $\left[\Mb_{d}\right]_{31}$ & $\left[\Mb_{d}\right]_{32}$ & $\left[\Mb_{d}\right]_{33}$ \\\hline
       Ch & 0 & 0 & 0 & 0 & 0 & 0 & 0 & 0 & -12\\\hline
       mag & 0 & 0 & 0 & 0 & 0 & 0 & 0 & 0 & -6$\kappa_n$\\\hline
       GT & 0 & 0 & 0 & 0 & 0 & 0 & 2 & 0 & 0\\\hline
       F & 0 & 0 & 0 & 0 & 0 & 0 & 0 & -6 & 0\\\hline
    \end{tabular}
    \caption{\label{tab:Md}Matrix elements of $\Mb_{d}$ for charge, magnetic, GT, and F form factors.}
\end{table}
Diagram (g) has the same form as diagram (e) but different c.c.~space matrix elements that we absorb into our expression for diagram (e).  The Coulomb correction to the generic three-nucleon form factor from diagrams (e) and (g) as well as well as the part of diagram (A) that only differs from diagram (e) by the c.c.~space matrix elements combine to give
\begin{align}
    &F_e(0)=-2\pi M_N^2\alpha\Gammab_0(p)\otimes_p\frac{1}{4\left(d(k)+d(p)\right)^2}\frac{1}{kp}\\\nonumber
    &\left[Q_0\left(\frac{p^2+k^2}{2kp}\right)-Q_0\left(\frac{k^2+p^2+4\left(d(k)+d(p)\right)^2}{2kp}\right)\right]\frac{1}{d(k)}\Mb_e\otimes_k\Gammab_0(k).
\end{align}
In the limit $q=p$ this expression possesses a logarithmic divergence.  In principle this divergence can be regulated by giving the photon a finite mass, instead we give the photon zero mass and use a subtraction technique to deal with the logarithmic divergence.  Table~\ref{tab:Me} gives the matrix elements of $\Mb_e$ for our form factors of interest. 
\begin{table}[hbt]
    \begin{tabular}{|c|c|c|c|c|c|c|c|c|c|}
        \hline
       &$\left[\Mb_{e}\right]_{11}$   & $\left[\Mb_{e}\right]_{12}$ & $\left[\Mb_{e}\right]_{13}$ & $\left[\Mb_{e}\right]_{21}$ & $\left[\Mb_{e}\right]_{22}$ & $\left[\Mb_{e}\right]_{23}$ & $\left[\Mb_{e}\right]_{31}$ & $\left[\Mb_{e}\right]_{32}$ & $\left[\Mb_{e}\right]_{33}$ \\\hline
       Ch & 8 & 0 & 0 & 0 & 24 & 0 & 0 & 0 & 0\\\hline
       mag & $\frac{4\kappa_p+8\kappa_n}{3}$ & $4(\kappa_p-\kappa_n)$ & 0 & $4(\kappa_p-\kappa_n)$ & $12\kappa_p$ & 0 & 0 & 0 & 0\\\hline
       GT & $\frac{2}{3}$ & 0 & -2 & 0 & 6 & 0 & 0 & 0 & 0\\\hline
       F & 2 & 0 & 0 & 0 & -6 & 6 & 0 & 0 & 0\\\hline
    \end{tabular}
    \caption{\label{tab:Me}Matrix elements of $\Mb_{e}$ for charge, magnetic, GT, and F form factors.}
\end{table}
Diagram (h) has the same form but different c.c.~space matrix elements from diagram (f) that we absorb into our expression for diagram (f).  The Coulomb correction to the generic three-nucleon form factor from diagrams (f) and (h) as well as the part of diagram (A) that only differs from diagram (f) by the c.c.~space matrix elements combine to give 
\begin{align}
    &F_f(0)=4\pi M_N^2\alpha\Gammab_0(p)\otimes_p\frac{1}{kp}\left\{\vphantom{\ln\left(\frac{p+k}{|p-k|}\right)}F_2[c,2(d(k)+d(p)),p,k]\right.\\\nonumber
    &\left.-\frac{1}{d(k)}\frac{1}{4(d(k)+d(p))^2-c^2}\left[Q_0\left(\frac{k^2+p^2+c^2}{2pk}\right)-Q_0\left(\frac{k^2+p^2+4(d(k)+d(p))^2}{2pk}\right)\right]\right.\\\nonumber
    &-2\frac{(d(k)-d(p))}{c^4}\left[\ln\left(\frac{p+k}{|p-k|}\right)-Q_0\left(\frac{k^2+p^2+c^2}{2pk}\right)-2pkc^2\frac{1}{(k^2+p^2+c^2)^2-4k^2p^2}\right]\\\nonumber
    &+\frac{3(k^2-p^2)}{c^4}\left(F_1\left[0,2(d(k)+d(p)),p,k\right]-F_1\left[c,2(d(k)+d(p)),p,k\right]\right.\\\nonumber
    &\left.\left.-c^2F_2\left[c,2(d(k)+d(p)),p,k\right]\right)\vphantom{\ln\left(\frac{p+k}{|p-k|}\right)}\right\}\Mb_f\otimes_k\Gammab_0(k),
\end{align}
where we define
\begin{equation}
    c=i\sqrt{3k^2+3p^2-2M_NE_B}.
\end{equation}
Table~\ref{tab:Mf} gives the matrix element of $\Mb_f$ for our form factors of interest.
\begin{table}[hbt]
    \begin{tabular}{|c|c|c|c|c|c|c|c|c|c|}
        \hline
       &$\left[\Mb_{f}\right]_{11}$   & $\left[\Mb_{f}\right]_{12}$ & $\left[\Mb_{f}\right]_{13}$ & $\left[\Mb_{f}\right]_{21}$ & $\left[\Mb_{f}\right]_{22}$ & $\left[\Mb_{f}\right]_{23}$ & $\left[\Mb_{f}\right]_{31}$ & $\left[\Mb_{f}\right]_{32}$ & $\left[\Mb_{f}\right]_{33}$ \\\hline
       Ch & -4 & 12 & 0 & 12 & 12 & 0 & 0 & 0 & 0\\\hline
       mag & $\frac{4\kappa_p-10\kappa_n}{3}$ & $(4\kappa_p+2\kappa_n)$ & 0 & $(4\kappa_p+2\kappa_n)$ & $6(2\kappa_p-\kappa_n)$ & 0 & 0 & 0 & 0\\\hline
       GT & $-\frac{1}{3}$ & 3 & 1 & 1 & 3 & -3 & 0 & 0 & 0\\\hline
       F & -1 & -3 & 3 & 3 & -3 & 3 & 0 & 0 & 0\\\hline
    \end{tabular}
    \caption{\label{tab:Mf}Matrix elements of $\Mb_{f}$ for charge, magnetic, GT, and F form factors.}
\end{table}
Finally the full $\mathcal{O}(\alpha)$ Coulomb correction to the generic three-nucleon form factor is given by the sum of these contributions yielding
\begin{equation}
F_{\alpha}(0)=F_{A+B+C}(0)+F_{a+b}(0)+F_c(0)+F_d(0)+F_e(0)+F_f(0).
\end{equation}

The Coulomb correction to the magnetic form factor can be simplified by rewriting it in terms of the Coulomb correction to the charge form factor yielding
\begin{align}
    \label{eq:Fmsimp}
    &F_\alpha^M(0)=\frac{1}{2}\kappa_nF_\alpha^C(0)+16\pi M_N^2\alpha\kappa_1\Gammab_0(p)\otimes_p\frac{1}{kp}\left\{\vphantom{\left[Q_0\left(\frac{p^2+k^2}{2kp}\right)-Q_0\left(\frac{k^2+p^2+4\left(d(k)+d(p)\right)^2}{2kp}\right)\right]}F_2[c,2(d(k)+d(p)),p,k]\right.\\\nonumber
    &-\frac{1}{d(k)}\frac{1}{4(d(k)+d(p))^2-c^2}\left[Q_0\left(\frac{k^2+p^2+c^2}{2pk}\right)-Q_0\left(\frac{k^2+p^2+4(d(k)+d(p))^2}{2pk}\right)\right]\\\nonumber
    &\left.-\frac{1}{8\left(d(k)+d(p)\right)^2}\left[Q_0\left(\frac{p^2+k^2}{2kp}\right)-Q_0\left(\frac{k^2+p^2+4\left(d(k)+d(p)\right)^2}{2kp}\right)\right]\frac{1}{d(k)}\right\}\\\nonumber
    &\left(\begin{array}{ccc}
    \frac{2}{3} & 2 & 0\\[-2mm]
    2 & 6 & 0\\[-2mm]
    0 & 0 & 0
    \end{array}\right)\otimes_k\Gammab_0(k)\\\nonumber
    &+4\pi M_N\kappa_1\Gammab_\alpha(p)\otimes_p\left\{\frac{\pi\delta(p-k)}{4k^2d(k)}-\frac{1}{(p^2+k^2-M_NE_B)^2-p^2k^2}\right\}\left(\begin{array}{ccc}
    \frac{2}{3} & 2 & 0\\[-2mm]
    2 & 6 & 0\\[-2mm]
    0 & 0 & 0
    \end{array}\right)\otimes_k\Gammab_0(k)\\\nonumber
    &+4\pi M_N\kappa_1\Gammab_0(E_B,p)\otimes_p\left\{\frac{\pi\delta(p-k)}{4k^2d(k)}-\frac{1}{(p^2+k^2-M_NE_B)^2-p^2k^2}\right\}\left(\begin{array}{ccc}
    \frac{2}{3} & 2 & 0\\[-2mm]
    2 & 6 & 0\\[-2mm]
    0 & 0 & 0
    \end{array}\right)\otimes_k\Gammab_\alpha(k).
\end{align}
By gauge symmetry it follows that $F_\alpha^C(0)=0$ and therefore the form of $F_\alpha^M(0)$ simplifies.  Likewise the GT form factor can be rewritten in terms of the Fermi form factor giving
\begin{equation}
    \label{eq:GTsimp}
    F_\alpha^{GT}(0)=F_\alpha^F(0)+\widetilde{F}_{A+B+C}^{GT}(0)+\widetilde{F}_{a+b}^{GT}(0)+\widetilde{F}_{c}^{GT}(0)+\widetilde{F}_{d}^{GT}(0)+\widetilde{F}_{e}^{GT}(0)+\widetilde{F}_{f}^{GT}(0),
\end{equation}
where the form factors $\widetilde{F}_{A+B+C}^{GT}(0)$, $\widetilde{F}_{a+b}^{GT}(0)$, $\widetilde{F}_{c}^{GT}(0)$, $\widetilde{F}_{d}^{GT}(0)$, $\widetilde{F}_{e}^{GT}(0)$, and $\widetilde{F}_{f}^{GT}(0)$ are the same as $F_{A+B+C}^{GT}(0)$, $F_{a+b}^{GT}(0)$, $F_{c}^{GT}(0)$, $F_{d}^{GT}(0)$, $F_{e}^{GT}(0)$, and $F_{f}^{GT}(0)$ except the matrices are now those in Tab.~\ref{tab:MGT} for each contribution to the form factor.
\begin{table}[hbt]
    \begin{tabular}{|c|c|c|c|c|c|c|c|c|c|}
        \hline
       &$\left[\widetilde{\Mb}_{x}\right]_{11}$   & $\left[\widetilde{\Mb}_{x}\right]_{12}$ & $\left[\widetilde{\Mb}_{x}\right]_{13}$ & $\left[\widetilde{\Mb}_{x}\right]_{21}$ & $\left[\widetilde{\Mb}_{x}\right]_{22}$ & $\left[\widetilde{\Mb}_{x}\right]_{23}$ & $\left[\widetilde{\Mb}_{x}\right]_{31}$ & $\left[\widetilde{\Mb}_{x}\right]_{32}$ & $\left[\widetilde{\Mb}_{x}\right]_{33}$ \\[1.5mm]\hline
       $\widetilde{\Mb}_1$ & $-\frac{2}{3}$ & 0 & -1 & 0 & 6 & -3 & -1 & -3 &0 \\\hline
       $\widetilde{\Mb}_2$ & $-\frac{4}{3}$ & 0 & -2 & 0 & 12 & -6 & -2 & -6 &0 \\\hline
       $\widetilde{\Mb}_{a+b}$ & -2 & -6 & 0 & 2 & 6 & 0 & 0 & 0 & 0\\\hline
       $\widetilde{\Mb}_c$ & 0 & 0 & 0 & 0 & 0 & 0 & -2 & -6 & 0\\\hline
       $\widetilde{\Mb}_d$ & 0 & 0 & 0 & 0 & 0 & 0 & 2 & 6 & 0\\\hline
       $\widetilde{\Mb}_e$ & $-\frac{4}{3}$ & 0 & -2 & 0 & 12 & -6 & 0 & 0 & 0\\\hline
       $\widetilde{\Mb}_f$ & $\frac{2}{3}$ & 6 & -2 & -2 & 6 & -6 & 0 & 0 & 0\\\hline
    \end{tabular}
    \caption{\label{tab:MGT}Matrix elements of all diagrams for GT form factor after adding F form factor, Eq.~\eqref{eq:GTsimp}.}
\end{table}
Although this form is not simpler for $F_\alpha^{GT}(0)$ it is essential for discerning the behavior of $F_\alpha^{GT}(0)$ under Wigner-SU(4) symmetry.

\section{\label{sec:wig}Wigner-SU(4) symmetry}

\subsection{Vertex Function}

To determine the properties of observables due to Wigner-SU(4) symmetry we need to transform to the Wigner-SU(4) basis.  The transformation has been carried out before in 2$\times$2~\cite{Griesshammer:2010nd,Bedaque:1999ve} and 3$\times$3~\cite{Vanasse:2014kxa} c.c.~space.  However, we will define the Wigner-SU(4) basis differently from Ref.~\cite{Vanasse:2014kxa}.  The LO vertex function, and the NLO and $\mathcal{O}(\alpha)$ corrections to the vertex function in the Wigner-SU(4) basis are given by 
\begin{equation}
    \left(\begin{array}{c}
    \G_{n,W\!s}(p)\\
    \G_{n,W\!as}(p)\\
    \G_{n,\emptyset}(p)\\
    \end{array}\right)=\Gb_{n,W}(p)=\Mb_W\Gb_n(p)\,\,,\,\left(\begin{array}{c}
    \G_{\alpha,W\!s}(p)\\
    \G_{\alpha,W\!as}(p)\\
    \G_{\alpha,\emptyset}(p)\\
    \end{array}\right)=\Gb_{\alpha,W}(p)=\Mb_W\Gb_\alpha(p),
\end{equation}
where $\G_{n,W\!s}(p)$ [$\G_{\alpha,W\!s}(p)$] is the ``Wigner symmetric" and $\G_{n,W\!as}(p)$ [$\G_{\alpha,W\!as}(p)$] the ``Wigner antisymmetric" component of $\Gb_{n,W}(p)$ [$\Gb_{\alpha,W}(p)$].  $\G_{0,\emptyset}(p)$ in our expansion can be shown to be zero while $\G_{\alpha,\emptyset}(p)$ will contain both Wigner symmetric and antisymmetric contributions. The matrix $\Mb_W$ is defined by 
\begin{equation}
    \Mb_W=\left(\begin{array}{rrr}
    1 & -3 & 0\\[-1mm]
    1 & 3 & 0\\[-1mm]
    0 &1 & -\frac{1}{2}
    \end{array}\right).
\end{equation}
By repeated use of the identity
\begin{equation}
    \label{eq:Wigid}
    \frac{1}{2}\mathbf{X}\Mb_W^T\Mb_W=\left(\begin{array}{ccc}
    1 & 0 & 0\\[-2mm]
    0 & 1 & 0 \\[-2mm]
    0 & 0 & 1
    \end{array}\right),
\end{equation}
where
\begin{equation}
    \mathbf{X}=\left(\begin{array}{ccc}
    1 & 0 & 0\\[-1mm]
    0 & \frac{1}{9} & \frac{2}{9}\\[-1mm]
    0 & \frac{2}{9} & \frac{76}{9}
    \end{array}\right),
\end{equation}
our expressions for the form factors and vertex functions can be rewritten in the Wigner-SU(4) basis.  In the Wigner-SU(4) basis we also define
\begin{equation}
    \Db_W(E_B,q)=\frac{1}{2}\Mb_W\mathbf{X}\Db(E_B,q)\Mb_W^T,
\end{equation}
which gives
\begin{equation}
    \Db_W(E_B,q)=\left(\begin{array}{ccc}
    D_+(E_B,q) & D_-(E_B,q) & 0\\
    D_-(E_B,q) & D_+(E_B,q) & 0\\
    0 & 0 & D_+(E_B,q)-D_-(E_B,q)
    \end{array}\right),
\end{equation}
where
\begin{equation}
    D_+(E_B,q)=\frac{1}{2}\left(D_t(E_B,q)+D_s(E_B,q)\right)\,\,,\,\, D_-(E_B,q)=\frac{1}{2}\left(D_t(E_B,q)-D_s(E_B,q)\right).
\end{equation}
Transforming the LO vertex function, Eq.~\eqref{eq:G}, to the Wigner-SU(4) basis gives
\begin{align}
\Gb_{0,W}(p)=\oneb_W+\mathbf{K}_W(E_B,p,q)\otimes_q\Db_W(E_B,q)\Gb_{0,W}(q),
\end{align}
where the inhomogeneous term, $\oneb_W$ in the Wigner-SU(4) basis is
\begin{equation}
    \oneb_W=\left(\begin{array}{r}
    2 \\
    0\\
    0
    \end{array}\right),
\end{equation}
and the kernel, $\mathbf{K}_W(E_B,p,q)$ in the Wigner-SU(4) basis is
\begin{equation}
\mathbf{K}_W(E_B,p,q)=-\frac{2\pi}{qp}Q_0\left(\frac{q^2+p^2-M_NE_B}{qp}\right)\left(\begin{array}{rrr}
    4 & 0 & 12\\
    0 & -2 & 0\\
    0 & 0 & -2
    \end{array}\right).
\end{equation}
From the integral equation for the LO vertex function in the Wigner-SU(4) basis one finds
\begin{equation}
    \G_{0,\emptyset}(p)=0,
\end{equation}
and our 3$\times$3 c.c.~space can be reduced to a 2$\times$2 c.c.~space similar to previous calculations~\cite{Vanasse:2016umz}.  In the Wigner-SU(4) limit, $\gamma_t=\gamma_s$, resulting in $D_-(E_B,p)=0$, and the Wigner symmetric and Wigner antisymmetric components of the vertex function decouple.  The Wigner antisymmetric component is found to be zero and only the Wigner symmetric component remains with its integral equation being equivalent to that of three-bosons~\cite{Bedaque:1998kg,Bedaque:1999ve}.  To account for the breaking of Wigner-SU(4) symmetry we define
\begin{equation}
    \gamma=\frac{1}{2}(\gamma_t+\gamma_s)\quad,\quad \delta=\frac{1}{2}(\gamma_t-\gamma_s).
\end{equation}
In the Wigner-SU(4) limit $\delta=0$ and thus $\delta$ accounts for the breaking of Wigner-SU(4) symmetry.  Following Ref.~\cite{Vanasse:2016umz} both the vertex function and dibaryon propagators can be expanded in powers of $\delta$ giving for the LO vertex function
\begin{equation}
    \G_{0,W\!s}(p)=\sum_{n=0}^\infty\delta^{2n}\G_{0,W\!s}^{(2n)}(p)\quad,\quad \G_{0,W\!as}(p)=\sum_{n=0}^\infty\delta^{2n+1}\G_{0,W\!as}^{(2n+1)}(p),
\end{equation}
and for the dibaryon propagators
\begin{equation}
    \label{eq:DWigexp}
    D_+(E_B,p)=\sum_{n=0}^\infty\delta^{2n}\left[D(E_B,p)\right]^{2n+1}\quad,\quad D_-(E_B,p)=-\sum_{n=0}^\infty\delta^{2n+1}\left[D(E_B,p)\right]^{2(n+1)},
\end{equation}
where\footnote{Note, the sign of $D(E_B,p)$ differs from Ref.~\cite{Vanasse:2016umz} resulting in sign differences for Eqs.~\eqref{eq:DWigexp},\eqref{eq:GtildeWs},\eqref{eq:GtildeWas}, and \eqref{eq:Gtildedef} with Ref.~\cite{Vanasse:2016umz}.}
\begin{equation}
    D(E_B,p)=\frac{1}{\gamma-\sqrt{\frac{3}{4}p^2-M_NE_B}}.
\end{equation}
Collecting like powers of $\delta$, integral equations for $\G_{0,W\!s}^{(2n)}(p)$ and $\G_{0,W\!as}^{(2n+1)}(p)$ can be constructed~\cite{Vanasse:2016umz}, yielding
\begin{align}
    \label{eq:GtildeWs}
&\widetilde{\G}_{0,W\!s}^{(2n)}(p)=2\delta_{0n}-D(E_B,p)\widetilde{\G}_{0,W\!as}^{(2n-1)}+M(E_B,p,q)\otimes_q \widetilde{\G}_{0,W\!s}^{(2n)}(q)
\end{align}
\begin{align}
\label{eq:GtildeWas}
\widetilde{\G}_{0,W\!as}^{(2n+1)}(p)=-D(E_B,p)\widetilde{\G}_{0,W\!s}^{(2n)}-\frac{1}{2}M(E_B,p,q)\otimes_q \widetilde{\G}_{0,W\!as}^{(2n+1)}(q),
\end{align}
where
\begin{equation}
    M(E_B,p,q)=-\frac{8\pi}{qp}Q_0\left(\frac{q^2+p^2-M_NE_B}{qp}\right)D(E_B,q). 
\end{equation}
$\widetilde{\G}_{0,W\!s}^{(2n)}(p)$ ]$\widetilde{\G}_{0,W\!as}^{(2n)}(p)$] is related to $\G_{0,W\!s}^{(2n)}(p)$ [$\G_{0,W\!as}^{(2n)}(p)$] by

\begin{align}
    \label{eq:Gtildedef}
    &\widetilde{\G}_{0,W\!s}^{(2n)}(p)=\G_{0,W\!s}^{(2n)}(p)-D(E_B,p)\widetilde{\G}_{0,W\!as}^{(2n-1)}(p)\\\nonumber
    &\widetilde{\G}_{0,W\!as}^{(2n+1)}(p)=\G_{0,W\!as}^{(2n+1)}(p)-D(E_B,p)\widetilde{\G}_{0,W\!s}^{(2n)}(p),
\end{align}
and is defined to make the integral equations simpler~\cite{Vanasse:2016umz}.  In calculations of form factors it is more convenient to use $\Gammab_0(p)$, which in the Wigner-SU(4) basis is expanded as
\begin{equation}
    \left(\begin{array}{c}
    \Gamma_{0,W\!s}(p)\\
    \Gamma_{0,W\!as}(p)\\
    0
    \end{array}\right)=\Gammab_{0,W}(p)=\mathcal{M}_W\Gammab_0(p)=\sum_{m=0}^{\infty}\delta^{m}\Gammab_{0,W}^{(m)}(p),
\end{equation}
where $\Gammab_{0,W}^{(m)}(p)$ is defined by
\begin{equation}
    \left(\begin{array}{c}
    \Gamma_{0,W\!s}^{(m)}(p)\\
    \Gamma_{0,W\!as}^{(m)}(p)\\
    0
    \end{array}\right)=\Gammab_{0,W}^{(m)}(p)=D(E_B,p)\Gbt_{0,W}(p)=D(E_B,p)\left(\begin{array}{c}
    \Gt_{0,W\!s}^{(m)}(p)\\
    \Gt_{0,W\!as}^{(m)}(p)\\
    0
    \end{array}\right).
\end{equation}
The leading scaling of the components of $\Gammab_{0,W}(p)$ in powers of $\delta$ is
\begin{equation}
    \Gamma_{0,W\!s}(p)\sim\mathcal{O}(\delta^0)\quad,\quad\Gamma_{0,W\!as}(p)\sim\mathcal{O}(\delta).
\end{equation}

\subsection{$\mathcal{O}(\alpha)$ correction to Vertex function}

By repeated use of Eq.~\eqref{eq:Wigid} the $\mathcal{O}(\alpha)$ Coulomb correction to the vertex function, Eq.~\eqref{eq:galpha}, can be written in the Wigner-SU(4) basis as
\begin{align}
    \label{eq:galphaW}
    \Gb_{\alpha,W}(p)=\mathbf{R}_{\alpha,W}(E_B,p)\Gammab_{0,W}(p)+&\mathbf{K}_{\alpha,W}(E_B,p,q)\Gammab_{0,W}(q)\\\nonumber
    &+\mathbf{K}_{0,W}(E_B,p,q)\otimes_q\Gb_{\alpha,W}(q).
\end{align}
The kernel $\mathbf{K}_{\alpha,W}(E_B,p,q)$ is again split into two contributions
\begin{equation}
    \mathbf{K}_{\alpha,W}(E_B,p,q)=\mathbf{K}_{\alpha,W}^{(SC)}(E_B,p,q)+\mathbf{K}_{\alpha,W}^{(C)}(E_B,p,q),
\end{equation}
where $\mathbf{K}_{\alpha,W}^{(SC)}(E_B,p,q)$ includes diagrams that mix strong and Coulomb interactions between dibaryon and nucleons while $\mathbf{K}_{\alpha,W}^{(C)}(E_B,p,q)$ includes diagrams that only have Coulomb interactions between the dibaryon and nucleon.  The matrix $\mathbf{K}_{\alpha,W}^{(SC)}(E_B,p,q)$ is given by
\begin{align}
    &\mathbf{K}_{\alpha,W}^{(SC)}(E_B,p,q)= C(E_B,p,q)\left(\begin{array}{rrr}
    2 & 2 & 0 \\[-2mm]
    0 & -2 & 0 \\[-2mm]
    -\frac{1}{3} &-\frac{2}{3} & 0
    \end{array}\right)\\\nonumber
    &-V_1(E_B,p,q)\left(\begin{array}{rrr}
    -2 & 2 & -12  \\[-2mm]
    0 & 0 & 0 \\[-2mm]
    \frac{1}{3} &-\frac{1}{3} &2
    \end{array}\right)+V_2(E_B,p,q)\left(\begin{array}{ccc}
    0 & 0 & 0 \\[-2mm]
    0 & 0 & 0 \\[-2mm]
    \frac{2}{3} &\frac{1}{3} & 0
    \end{array}\right),
\end{align}
while $\mathbf{K}_{\alpha,W}^{(C)}(E_B,p,q)$ is
\begin{align}
    &\mathbf{K}_{\alpha,W}^{(C)}(E_B,p,q)= B(E_B,p,q)\left(\begin{array}{rrr}
    1 & 0 & 0\\[-2mm]
    0 & 1 & 0 \\[-2mm]
    -\frac{1}{6} &\frac{1}{6} &0
    \end{array}\right).
\end{align}
$\mathbf{R}_{\alpha,W}(E_B,p)$ in the Wigner-SU(4) basis is given by 
\begin{equation}
    \label{eq:RalphaW}
    \mathbf{R}_{\alpha,W}(E_B,p)=\left[-\frac{1}{a_C}+\gamma_s+\alpha M_N\gamma_E+\alpha M_N\ln\left(\frac{\alpha M_N}{2d(p)}\right)\right]\left(\begin{array}{ccc}
    0 & 0 & 0\\[-2mm]
    0 & 0 & 0\\[-2mm]
    \frac{1}{6} & -\frac{1}{6} & 1
    \end{array}\right).
\end{equation}
The $\mathcal{O}(\alpha)$ Coulomb correction to the vertex function can be expanded in powers of $\delta$ giving
\begin{equation}
    \G_{\alpha,W\!s}(p)=\sum_{n=0}^{\infty}\delta^{n}\G_{\alpha,W\!s}^{(n)}(p)\,\,,\,\,\G_{\alpha,W\!as}(p)=\sum_{n=1}^{\infty}\delta^{n}\G_{\alpha,W\!as}^{(n)}(p)\,\,,\,\,\G_{\alpha,\emptyset}(p)=\sum_{n=0}^{\infty}\delta^{n}\G_{\alpha,\emptyset}^{(n)}(p).
\end{equation}
Expanding the LO vertex function and the $\mathcal{O}(\alpha)$ Coulomb correction to the vertex function and collecting like powers of $\delta$ gives integral equations for the order-by-order $\delta$ corrections to the $\mathcal{O}(\alpha)$ Coulomb correction to the vertex function.  To simplify these integral equations we define
\begin{align}
    &\widetilde{\G}_{\alpha,W\!s}^{(2n)}(p)=\G_{\alpha,W\!s}^{(2n)}(p)-D(E_B,p)\widetilde{\G}_{\alpha,W\!as}^{(2n-1)}(p)\\\nonumber
    &\widetilde{\G}_{\alpha,W\!as}^{(2n+1)}(p)=\G_{\alpha,W\!as}^{(2n+1)}(p)-D(E_B,p)\widetilde{\G}_{\alpha,W\!s}^{(2n)}(p)\\\nonumber
    &\widetilde{\G}_{\alpha,\emptyset}^{(n)}(p)=\G_{\alpha,\emptyset}^{(n)}(p)+D(E_B,p)\widetilde{\G}_{\alpha,\emptyset}^{(n-1)}(p).
\end{align}
Using this redefinition of the vertex function we find for even orders of $\delta$ the integral equations 
\begin{align}
&\widetilde{\G}_{\alpha,W\!s}^{(2n)}(p)=(2(C(E_B,p,q)+V_1(E_B,p,q))+B(E_B,p,q))\otimes_q D(E_B,q)\Gt_{0,W\!s}^{(2n)}(q)\\\nonumber
&\hspace{1cm}-D(E_B,p)\widetilde{\G}_{\alpha,W\!as}^{(2n-1)}(p)+M(E_B,p,q)\otimes_q \widetilde{\G}_{\alpha,W\!s}^{(2n)}(q)+3M(E_B,p,q)\otimes_q \widetilde{\G}_{\alpha,\emptyset}^{(2n)}(q),
\end{align}
\begin{align}
\Gt_{\alpha,W\!as}^{(2n)}(p)=-D(E_B,p)\Gt_{\alpha,W\!s}^{(2n-1)}(p)-\frac{1}{2}M(E_B,p,q)\otimes_q\widetilde{\G}_{\alpha,W\!as}^{(2n)}(q),
\end{align}
and
\begin{align}
&\Gt_{\alpha,\emptyset}^{(2n)}(p)=\frac{1}{3}(2V_2(E_B,p,q)-V_1(E_B,p,q)-C(E_B,p,q)\\\nonumber
&\hspace{1cm}-\frac{1}{2}B(E_B,p,q))\otimes_q D(E_B,q)\Gt_{0,W\!s}^{(2n)}(q)+R_\alpha(E_B,p)\Gt_{0,W\!s}^{(2n)}(p)\\\nonumber
&\hspace{1cm}+\frac{1}{6}D(E_B,p)\Gt_{0,W\!as}^{(2n-1)}(p)+D(E_B,p)\Gt_{\alpha,\emptyset}^{(2n-1)}(p)-\frac{1}{2}M(E_B,p,q)\otimes_q\Gt_{\alpha,\emptyset}^{(2n)}(q).
\end{align}
$R_{\alpha}(E_B,p)$ is defined by
\begin{equation}
    R_{\alpha}(E_B,p)=\frac{1}{6}\left[-\frac{1}{a_C}+\gamma+\alpha M_N\gamma_E+\alpha M_N\ln\left(\frac{\alpha M_N}{2d(p)}\right)\right]D(E_B,p).
\end{equation}
For odd orders of $\delta$ we get the integral equations
\begin{align}
&\widetilde{\G}_{\alpha,W\!s}^{(2n+1)}(p)=2 (C(E_B,p,q)-V_1(E_B,p,q))\otimes_q D(E_B,q)\Gt_{0,W\!as}^{(2n+1)}(q)\\\nonumber
&\hspace{1cm}-D(E_B,p)\widetilde{\G}_{\alpha,W\!as}^{(2n)}(p)+M(E_B,p,q)\otimes_q \widetilde{\G}_{\alpha,W\!s}^{(2n+1)}(q)+3M(E_B,p,q)\otimes_q \widetilde{\G}_{\alpha,\emptyset}^{(2n+1)}(q),
\end{align}
\begin{align}
&\Gt_{\alpha,W\!as}^{(2n+1)}(p)=(B(E_B,p,q)-2C(E_B,p,q))\otimes_q D(E_B,q)\Gt_{0,W\!as}^{(2n+1)}(q)\\\nonumber
&\hspace{1cm}-D(E_B,p)\Gt_{\alpha,W\!s}^{(2n)}(p)-\frac{1}{2}M(E_B,p,q)\otimes_q\widetilde{\G}_{\alpha,W\!as}^{(2n+1)}(q),
\end{align}
and
\begin{align}
&\Gt_{\alpha,\emptyset}^{(2n+1)}(p)=\frac{1}{3}(V_2(E_B,p,q)+V_1(E_B,p,q)-2C(E_B,p,q)\\\nonumber
&\hspace{1cm}+\frac{1}{2}B(E_B,p,q))\otimes_q D(E_B,q)\Gt_{0,W\!as}^{(2n+1)}(q)-R_\alpha(E_B,p)\Gt_{0,W\!as}^{(2n+1)}(p)\\\nonumber
&\hspace{1cm}-\frac{1}{6}D(E_B,p)\Gt_{0,W\!s}^{(2n)}(p)+D(E_B,p)\Gt_{\alpha,\emptyset}^{(2n)}(p)-\frac{1}{2}M(E_B,p,q)\otimes_q\Gt_{\alpha,\emptyset}^{(2n+1)}(q).
\end{align}
The $\frac{1}{6}D(E_B,p)\Gt_{\alpha,W\!as}^{(2n-1)}(p)$ [$-\frac{1}{6}D(E_B,p)\Gt_{\alpha,W\!s}^{(2n)}(p)$] term appearing in the inhomogeneous term of $\Gt_{\alpha,\emptyset}^{(2n)}(p)$ [$\Gt_{\alpha,\emptyset}^{(2n+1)}(p)$] comes from expanding the $\delta$ in the $\gamma_s$ of $\mathbf{R}_{\alpha,W}(E_B,p)$ in Eq.~\eqref{eq:RalphaW}.  In calculations of form factors it is more convenient to use $\Gammab_\alpha(p)$, which in the Wigner-SU(4) basis is expanded as
\begin{equation}
    \left(\begin{array}{c}
    \Gamma_{\alpha,W\!s}(p)\\
    \Gamma_{\alpha,W\!as}(p)\\
    \Gamma_{\alpha,\emptyset}(p)
    \end{array}\right)=\Gammab_{\alpha,W}(p)=\mathcal{M}_W\Gammab_\alpha(p)=\sum_{m=0}^{\infty}\delta^{m}\Gammab_{\alpha,W}^{(m)}(p),
\end{equation}
where $\Gammab_{\alpha,W}^{(m)}(p)$ is defined by
\begin{equation}
    \left(\begin{array}{c}
    \Gamma_{\alpha,W\!s}^{(m)}(p)\\
    \Gamma_{\alpha,W\!as}^{(m)}(p)\\
    \Gamma_{\alpha,\emptyset}^{(m)}(p)
    \end{array}\right)=\Gammab_{\alpha,W}^{(m)}(p)=D(E_B,p)\Gbt_{\alpha,W}^{(m)}(p)=D(E_B,p)\left(\begin{array}{c}
    \Gt_{\alpha,W\!s}^{(m)}(p)\\
    \Gt_{\alpha,W\!as}^{(m)}(p)\\
    \Gt_{\alpha,\emptyset}^{(m)}(p)
    \end{array}\right).
\end{equation}
The leading scaling of the components of $\Gammab_{\alpha,W}(p)$ in powers of $\delta$ is
\begin{equation}
    \Gamma_{\alpha,W\!s}(p)\sim\mathcal{O}(\delta^0)\quad,\quad\Gamma_{\alpha,W\!as}(p)\sim\mathcal{O}(\delta)\quad,\quad\Gamma_{\alpha,\emptyset}(p)\sim\mathcal{O}(\delta^0).
\end{equation}

\subsection{LO helium-3 magnetic moment and GT form factor}

By repeated use of Eq.~\eqref{eq:Wigid} the LO magnetic moment, Eq.~\eqref{eq:LOF}, can be rewritten in the Wigner-SU(4) basis.  Also taking the part of $F_M^{\jjvHe}(0)$ that goes like $F_C^{\jjvHe}(0)$ at LO and noting that at LO $F_C^{\jjvHe}(0)=2$ the LO $\jjvHe$ magnetic moment is given by
\begin{align}
    &\mu_0^{\jjvHe}=\kappa_n \\\nonumber
    &\hspace{.5cm}+\frac{8}{3}\pi M_N\kappa_1\Gamma_{0,W\!as}(p)\otimes_p\left\{\frac{\pi}{4}\frac{\delta(k-p)}{k^2d(k)}-\frac{1}{(k^2+p^2-M_NE_B)^2-k^2p^2}\right\}\otimes_k\Gamma_{0,W\!as}(k).
\end{align}
Carrying out a similar exercise for the LO GT matrix element yields
\begin{align}
    \label{eq:GTLO}
    &\GT_0=1 \\\nonumber
    &\hspace{.5cm}-\frac{8}{3}\pi M_N\Gamma_{0,W\!as}(p)\otimes_p\left\{\frac{\pi}{4}\frac{\delta(k-p)}{k^2d(k)}-\frac{1}{(k^2+p^2-M_NE_B)^2-k^2p^2}\right\}\otimes_k\Gamma_{0,W\!as}(k).
\end{align}
To expand these observables in powers of $\delta$ we replace $\Gamma_{0,W\!as}(p)$ with its $\delta$ expansion.  Since $\Gamma_{0,W\!as}(p)$ only has odd powers of $\delta$  we find that $\mu_0^{\jjvHe}$ and $\GT_0$ only have even powers of $\delta$.  It is also worth noting that the $\delta$ corrections to $\mu_{0}^{\jjvHe}$ and $\GT_0$ are exactly the same up to a factor of $-\kappa_1$.

\subsection{$\mathcal{O}(\alpha)$ correction to helium-3 magnetic moment}
The $\mathcal{O}{(\alpha)}$ correction to $\mu^{\jjvHe}$ given by Eq.~\eqref{eq:Fmsimp} can be rewritten in the Wigner-SU(4) basis by repeated use of the identity Eq.~\eqref{eq:Wigid} yielding

\begin{align}
    &F_\alpha^M(0)=16\pi M_N^2\alpha\frac{2}{3}\kappa_1\Gamma_{0,W\!as}(p)\otimes_p\frac{1}{kp}\left\{F_2[c,2(d(k)+d(p)),p,k]\right.\\\nonumber
    &-\frac{1}{d(k)}\frac{1}{4(d(k)+d(p))^2-c^2}\left[Q_0\left(\frac{k^2+p^2+c^2}{2pk}\right)-Q_0\left(\frac{k^2+p^2+4(d(k)+d(p))^2}{2pk}\right)\right]\\\nonumber
    &\left.-\frac{1}{8\left(d(k)+d(p)\right)^2}\left[Q_0\left(\frac{p^2+k^2}{2kp}\right)-Q_0\left(\frac{k^2+p^2+4\left(d(k)+d(p)\right)^2}{2kp}\right)\right]\frac{1}{d(k)}\right\}\otimes_k\Gamma_{0,W\!as}(k)\\\nonumber
    &+4\pi M_N\frac{2}{3}\kappa_1\Gamma_{\alpha,W\!as}(p)\otimes_p\left\{\frac{\pi\delta(p-k)}{4k^2d(k)}-\frac{1}{(p^2+k^2-M_NE_B)^2-p^2k^2}\right\}\otimes_k\Gamma_{0,W\!as}(k)\\\nonumber
    &+4\pi M_N\frac{2}{3}\kappa_1\Gamma_{0,W\!as}(E_B,p)\otimes_p\left\{\frac{\pi\delta(p-k)}{4k^2d(k)}-\frac{1}{(p^2+k^2-M_NE_B)^2-p^2k^2}\right\}\otimes_k\Gamma_{\alpha,W\!as}(k),
\end{align}
where we have used gauge symmetry, which gives $F_\alpha^{(C)}(0)=0$.  Noting the leading scalings $\Gamma_{0,W\!as}(p)\sim\Gamma_{\alpha,W\!as}\sim\mathcal{O}(\delta)$ we find that the $\mathcal{O}(\alpha)$ correction to $\mu^{\jjvHe}$ has the leading scaling of $\mathcal{O}(\delta^2)$.  To find the $\delta$ expansion of this observable we replace $\Gamma_{0,W\!as}(p)$ and $\Gamma_{\alpha,W\!as}(p)$ with their $\delta$ expansions and then collect like powers of $\delta$.

\subsection{$\mathcal{O}(\alpha)$ correction to $GT$ matrix element}

The $\mathcal{O}(\alpha)$ correction to the GT matrix element is given by Eq.~\eqref{eq:GTsimp} and the corresponding matrices listed in Tab.~\ref{tab:MGT}.  These matrices can be rewritten in the Wigner-SU(4) basis by repeated use of Eq.~\eqref{eq:Wigid} giving the matrices in Tab.~\ref{tab:MGTW}
\begin{table}[hbt]
    \begin{tabular}{|c|c|c|c|c|c|c|c|c|c|}
        \hline
       &$\left[\Mb_{x}\right]_{11}$   & $\left[\Mb_{x}\right]_{12}$ & $\left[\Mb_{x}\right]_{13}$ & $\left[\Mb_{x}\right]_{21}$ & $\left[\Mb_{x}\right]_{22}$ & $\left[\Mb_{x}\right]_{23}$ & $\left[\Mb_{x}\right]_{31}$ & $\left[\Mb_{x}\right]_{32}$ & $\left[\Mb_{x}\right]_{33}$ \\\hline
       $\Mb_1$ & 0 & 0 & 0 & 0 & -2/3 & 2 & 0 & 2 &0 \\\hline
       $\Mb_2$ & 0 & 0 & 0 & 0 & -4/3 & 4 & 0 & 4 &0 \\\hline
       $\Mb_{a+b}$ & 0 & -4/3 & 0 & 0 & -2/3 & 0 & 0 & 0 & 0\\\hline
       $\Mb_c$ & 0 & 2/3 & 0 & 0 & -2/3 & 0 & 0 & 4 & 0\\\hline
       $\Mb_d$ & 0 & -2/3 & 0 & 0 & 2/3 & 0 & 0 & -4 & 0\\\hline
       $\Mb_e$ & 0 & -2/3 & 0 & 0 & -2/3 & 4 & 0 & 0 & 0\\\hline
       $\Mb_f$ & 0 & 2/3 & 0 & 0 & 0 & 4 & 0 & 0 & 0\\\hline
    \end{tabular}
    \caption{\label{tab:MGTW} Matrix elements of all diagrams for $F_{\alpha}^{GT}(0)$ in the Wigner-SU(4) basis}
\end{table}
These matrices have no component connecting Wigner symmetric to Wigner symmetric but do have a non-zero component connecting Wigner symmetric to Wigner antisymmetric.  Thus the GT matrix element has the naive leading scaling of $\mathcal{O}(\delta)$ in the Wigner-SU(4) expansion.  To obtain the $\delta$ expansion of this observable we again replace all vertex functions with their $\delta$ expansions and collect all like powers of $\delta$.

\section{\label{sec:results} Results}

\subsection{Helium-3 magnetic moment and GT matrix element}

The magnetic form factor at $Q^2=0$ gives the three-nucleon magnetic moment. Defining
\begin{equation}
    \mu^{\jjvHe}_0=F_0^M(0)\quad,\quad\mu^{\jjvHe}_1=F_1^M(0)\quad,\quad \mu^{\jjvHe}_\alpha=F_\alpha^M(0),
\end{equation}
we find the the $\jjvHe$ magnetic moment up to NLO and $\mathcal{O}(\alpha)$ is given by
\begin{equation}
    \mu^{\jjvHe}=\mu^{\jjvHe}_0+\mu^{\jjvHe}_1+\mu^{\jjvHe}_\alpha+\left(\frac{d}{dE}\mu^{\jjvHe}_0\Big{|}_{E=E_B}\right)B_1^{(\alpha)}.
\end{equation}
Note, the last term comes from the $\mathcal{O}(\alpha)$ correction to the three-nucleon binding energy that splits the $\jjvH$ and $\jjvHe$ binding energies.  We find a value of $B_1^{(\alpha)}=0.875(302)$~MeV that agrees within uncertainty with the experimental value of $B_1^{(\alpha)}=0.764$~MeV.  Our value is also close to the perturbative calculation of Ref.~\cite{Konig:2015aka} that found $B_1^{(\alpha)}=0.86(17)$~MeV.  Our results differ from Ref.~\cite{Konig:2015aka} as they expanded about the unitary limit in the $^{1}S_0$ channel and we expanded about the physical virtual bound state pole in the ${}^1S_0$ channel. 

The GT form factor at $Q^2=0$ gives the GT matrix element for $\jjvH$ $\beta$-decay.  Defining
\begin{equation}
    \GT_0=F_0^{GT}(0)\quad,\quad\GT_1=F_1^{GT}(0)\quad,\quad\GT_\alpha=F_\alpha^{GT}(0),
\end{equation}
the GT matrix element up to NLO and $\mathcal{O}(\alpha)$ is given by
\begin{equation}
    \GT=\GT_0+\GT_1+\GT_\alpha+\frac{1}{2}\left(\frac{d}{dE}\GT_0\Big{|}_{E=E_B}\right)B_1^{(\alpha)}.
\end{equation}
The last term again comes from the $\mathcal{O}(\alpha)$ correction to the three-nucleon binding energy while the additional factor of $1/2$ is because before the weak current the nuclear state is $\jjvH$ while only after the weak current is the nuclear state $\jjvHe$ and Coulomb corrections must be included. A summary of $\mu^{\jjvHe}$ and $\GT$ up to NLO and $\mathcal{O}(\alpha)$ can be found in Tab.~\ref{tab:results}.  
\begin{table}[hbt]
    \centering
    \begin{tabular}{|c|c|c|c|}
        \hline
         & Results & Experimental value \\\hline
         $\mu^{\jjvHe}$ &  $-1.8675\times\Big{(}\,\underbrace{\vphantom{0.9890^*+0.1503^*\times L_1}1}_{\mathrm{LO}}+\underbrace{0.9890+\left(29.66~\mathrm{MeV}\right)\times L_1}_{\mathrm{NLO}}-\underbrace{\vphantom{0.9890^*+0.1503^*\times L_1}0.0018}_{\mathcal{O}(\alpha)}\,\Big{)}$ & -2.127 \\ 
         $\GT$ & $0.9807\times\Big{(}\,\underbrace{\vphantom{0.7925+56.4833^*\times l_{1,A}}1}_{\mathrm{LO}}+\underbrace{0.7925-0.09548\times l_{1,A}}_{\mathrm{NLO}}-\underbrace{\vphantom{0.7925-0.0477\times l_{1,A}}0.00076}_{\mathcal{O}(\alpha)}\,\Big{)}$ & 
         0.9511\,\,\cite{Baroni:2016xll} \\\hline
    \end{tabular}
    \caption{LO, NLO and $\mathcal{O}(\alpha)$ correction to $\mu^{\jjvHe}$ and $\GT$ compared to experiment.  Perturbative corrections are normalized by the LO values and for the NLO correction the contribution from $L_{1,A}$ ($\ell_{1,A}$) for $\mu^{\jjvHe}$ ($\GT$) is separated out.}
    \label{tab:results}
\end{table}
The NLO correction to $\mu_{\jjvHe}$ ($\GT$) has been split into a contribution not from $L_1$ ($\ell_{1A}$) and from $L_1$ ($\ell_{1A}$). 
The value of $\mu^{\jjvHe}_1$ is determined by fitting $L_{1}$.  As in Ref.~\cite{Vanasse:2017kgh} we consider $L_1$ fit to the cold neutron-proton capture cross-section ($\sigma_{np}$), the $\jjvH$ magnetic moment ($\mu^{\jjvH}$), or simultaneously to $\sigma_{np}$ and $\mu^{\jjvH}$. The corresponding values for each fitting are presented in table \ref{tab:L1fit} with and without Coulomb corrections.
\begin{table}[hbt]
    \centering
    \begin{tabular}{|c|c|c|c|}
        \hline
        $L_1$-fit & $L_1$(fm )& $\mu^{\jjvHe}_{\text{LO+NLO}}$ & $\mu^{\jjvHe}_{\text{LO+NLO+$\alpha$}}$  \\\hline
         $\sigma_{np}$ & -6.90 & -1.777(212) & -1.774(212)  \\ 
         $\mu^{\jjvH}$ & -5.62 & -2.136(255) & -2.133(254) \\
         $\sigma_{np}$ and $\mu^{\jjvH}$ & -5.83 & -2.078(248) & -2.075(247) \\\hline
         Experimental value & N/A & -2.127 & -2.127 \\\hline
    \end{tabular}
    \caption{$\mu^{\jjvHe}$ in units of nuclear magnetons for different fits of the LEC $L_1$. LO, LO+NLO, and LO+NLO+$\mathcal{O}(\alpha)$ values of $\mu^{\jjvHe}$ are shown as well as the different values of $L_1$ for different fits.}
    \label{tab:L1fit}
\end{table}
Errors shown in Tab.~\ref{tab:L1fit} are a naive \EFT error estimate coming from $Q/\Lambda_{\not{\pi}}\sim(Z_t-1)/2\approx 0.35$.

\subsection{Value of $l_{1,A}$ and $pp$-fusion rate}
The $pp$ fusion rate depends on the matrix element (as shown in Ref.~\cite{Kong:2000px})
\begin{equation}
 \left|\langle d;j\left|A_{k}^{-}\right|pp\rangle\right|=g_{A}C_{\eta}\sqrt{\frac{32\pi}{\gamma_{t}^{3}}}\Lambda(p)\delta_{k}^{j},
 \end{equation}
where $\Lambda(p)$, the reduced matrix element, at threshold is given by~\cite{Ando:2008va}\footnote{Note, the expression for $\Lambda(0)$ in our formalism can be derived from Ref.~\cite{Ando:2008va} by reparametrizing their Lagrangian.}
\begin{align}
\Lambda(0)&=\frac{1}{2}(1+Z_{t})\left\{e^{\chi}-\gamma_{t}a_{C}[1-\chi e^{\chi}\Gamma(0,\chi)]\right\}
+\frac{\gamma_{t}^{2}a_{C}}{M_Ng_A}l_{1,A}.
\end{align}
$\Gamma(0,\chi)$ is the incomplete gamma function and $\chi=(\alpha M_N)/\gamma_t$. $\Lambda(0)$ depends on the same LEC, $\ell_{1,A}$, as $\jjvH$ $\beta$-decay, thus by fitting $\ell_{1,A}$ to $\jjvH$ $\beta$-decay we can make a prediction for $\Lambda(0)$ at NLO.  The tritium half life in terms of the GT matrix element and F matrix element is given by
\begin{equation}
    \label{eq:Thalf}
    \frac{(1+\delta_R)f_V}{K/G_V^2}t_{1/2}=\frac{1}{\F^2+(f_A/f_V)g_A^2\GT^2}.
\end{equation}
Ref.~\cite{Baroni:2016xll} extracted a value of $\GT_{exp}=0.9511\pm0.0013$ from experiment.  Values of parameters in Eq.~\eqref{eq:Thalf} can be found in Ref.~\cite{Baroni:2016xll}.  To determine the Coulomb corrections to $l_{1,A}$ we split it up into
\begin{equation}
    l_{1,A}=l_{1,A}^{(0)}+l_{1,A}^{(\alpha)},
\end{equation}
where $l_{1,A}^{(0)}$ [$l_{1,A}^{(\alpha)}$] is the $\mathcal{O}(\alpha^0)$ [$\mathcal{O}(\alpha)$] contribution to $l_{1,A}$.  First ignoring Coulomb corrections to $\GT$ and matching to $\GT_{exp}$ we extract for $l_{1,A}^{(0)}$
\begin{equation}
    l_{1,A}^{(0)}=8.615\pm 2.98,
\end{equation}
where the error is a naive \EFT error estimate.  Ensuring that the Coulomb corrections to $\GT$ do not change the fit to $\GT_{exp}$ we find for $l_{1,A}^{(\alpha)}$
\begin{equation}
    l_{1,A}^{(\alpha)}=-0.007908(2731),
\end{equation}
where again the error is a naive \EFT error estimate.  Using our values for $l_{1,A}$ we find a value of $\Lambda(0)=2.776(331)$ which agrees within uncertainty with the phenomenological value of $\Lambda(0)=2.65(1)$~\cite{Adelberger:2010qa}.  Note, we show $\Lambda(0)$ to four digits as this is the first digit where the Coulomb correction to $l_{1,A}$ changes the value.  It is worth noting that our calculation is still not fully consistent due to the fact that Coulomb corrections are treated perturbatively in $\jjvH$ $\beta$-decay, to determine $l_{1,A}$, and then Coulomb is treated nonperturbatively in $pp$ fusion.  However, since the $\mathcal{O}(\alpha)$ Coulomb correction to $\jjvH$ $\beta$-decay is very small we expect this to have little effect to the order in \EFT that we are working.

\subsection{Wigner-SU(4) expansion}

Naively Coulomb corrections should scale as $\alpha M_N/p^*\approx 8\%$ where the three-body binding momentum $p^*\sim 88.5$~MeV.  However, the Coulomb correction to $\mu^{\jjvHe}$ and $\GT$, as seen in Tab.~\ref{tab:results} are considerably smaller than this naive estimate.  As noted earlier, the Coulomb correction to $\mu^{\jjvHe}$ is suppressed by $\mathcal{O}(\delta^2)$ in the Wigner-SU(4) expansion.  Assuming as in Ref.~\cite{Vanasse:2016umz} that $\mathcal{O}(\delta)\sim\mathcal{O}(\rho^2)$ where $\rho\sim Q/\Lambda_{\not{\pi}}$ we find for the the Coulomb correction to $\mu^{\jjvHe}$ an estimate of $\rho^4\alpha M_N/p^*\approx 0.11\%$ which is quite close to the size of the correction observed for $\mu^{\jjvHe}$ in Tab.~\ref{tab:results}.  The $\GT$ matrix element is only suppressed by $\mathcal{O}(\delta)$ in the Wigner-SU(4) expansion leading to the estimate $\rho^2\alpha M_N/p^*\approx 0.92\%$, which is noticeably larger than the observed $0.075\%$ in Tab.~\ref{tab:results}.  In order to understand this better we analytically carried out a Wigner-SU(4) expansion for the LO and $\mathcal{O}(\alpha)$ correction for $\mu^{\jjvHe}$ and $\GT$ shown in Tab.~\ref{tab:Wig-exp}.
\begin{table}[hbt]
    \centering
    \begin{tabular}{|c|cccc|}
    \hline
    & $\mu_0^{\jjvHe}$ & $\GT_0$ & $\mu_{\alpha}^{\jjvHe}$ & $\GT_{\alpha}$\\\hline
    $\mathcal{O}(\delta^0)$ & -1.91304 & 1 & 0 & 0\\
    $\mathcal{O}(\delta)$ & 0 & 0 & 0 & -8.39$\times 10^{-3}$\\
    $\mathcal{O}(\delta^2)$ & 4.13$\times 10^{-2}$ & -1.76$\times 10^{-2}$ & 1.45$\times 10^{-2}$ & 9.46$\times 10^{-3}$\\
    $\mathcal{O}(\delta^3)$ & 0 & 0 & -1.56$\times 10^{-2}$ & -2.77$\times 10^{-3}$\\
    $\mathcal{O}(\delta^4)$ & 3.78$\times 10^{-3}$ & -1.61$\times 10^{-3}$ & 6.61$\times 10^{-3}$ & 1.28$\times 10^{-3}$\\
    $\mathcal{O}(\delta^5)$ & 0 & 0 & -3.13$\times 10^{-3}$ & -4.66$\times 10^{-4}$\\
    $\mathcal{O}(\delta^6)$ & 3.79$\times 10^{-4}$ & -1.61$\times 10^{-4}$ & 1.27$\times 10^{-3}$ & 1.86$\times 10^{-4}$\\\hline
    $\displaystyle\sum_{n=0}^6\mathcal{O}(\delta^n)$ & -1.86757 & 0.98067 & 3.73$\times 10^{-3}$ & -6.88$\times 10^{-4}$\\
    $\displaystyle\sum_{n=0}^\infty\mathcal{O}(\delta^n)$ & -1.86752 & 0.98065 & 3.35$\times 10^{-3}$ & -7.40$\times 10^{-4}$\\\hline
    \end{tabular}
    \caption{Order by order $\delta$ corrections to $\GT_0$, $\mu_0^{\jjvHe}$, $\GT_{\alpha}$, and $\mu_{\alpha}^{\jjvHe}$. Also included are the values when Wigner-SU(4) is treated nonperturbatively (included in Tab.~\ref{tab:results}) and the sum of all $\delta$ corrections up to $\mathcal{O}(\delta^6)$.}
    \label{tab:Wig-exp}
\end{table}

$\GT_0$ at $\mathcal{O}(\delta^0)$ is related to the F form factor that by gauge symmetry has a value of exactly one at $Q^2=0$. $\mu_{0}^{\jjvHe}$ at $\mathcal{O}(\delta^0)$ is $\kappa_n$ since it can be related to $\frac{1}{2}\kappa_n$ times the charge form factor which again by gauge symmetry has a value of two at $Q^2=0$.  In the limit $\delta=0$ the three-nucleon wavefunction is spatially symmetric and $\mu_0^{\jjvHe}$ is given by the magnetic moment of the unpaired nucleon in what is known as the Schmidt limit~\cite{Schmidt1937}. The higher order corrections for $\GT_0$ and $\mu^{\jjvHe}$ only occur for even powers of $\delta$.  Summing all Wigner-SU(4) corrections to $\mathcal{O}(\delta^6)$ we find that $\mu_{0}^{\jjvHe}$ and $\GT_0$ converge nicely to the values of $\mu_{0}^{\jjvHe}$ and $\GT_0$ when Wigner-SU(4) corrections are treated nonperturbatively.

The first $\delta$ order of $\GT_{\alpha}$ [$\mu_{\alpha}^{\jjvHe}$], starting at $\mathcal{O}(\delta)$ [$\mathcal{O}(\delta^2)$], is nearly 11 [4.3] times the value of $\GT_{\alpha}$ [$\mu_{\alpha}^{\jjvHe}$] when Wigner-SU(4) corrections are treated nonperturbatively, which is not perturbatively close.  Moreover, the $\mathcal{O}(\delta^2)$ [$\mathcal{O}(\delta^3)$] contribution to $\GT_{\alpha}$ [$\mu_{\alpha}^{\jjvHe}$] is nearly the same size as the $\mathcal{O}(\delta)$ [$\mathcal{O}(\delta^2)$] term but opposite in sign.  Indeed this pattern continues but to a lesser extent, for the subsequent orders of $\delta$.  The sign alternates for each subsequent order for both $\GT_{\alpha}$ and $\mu_{\alpha}^{\jjvHe}$.  Summing all the Wigner-SU(4) corrections to $\mathcal{O}(\delta^6)$ for $\mu_{\alpha}^{\jjvHe}$ and $\GT_{\alpha}$ we find they are not especially close to the value of $\mu_{\alpha}^{\jjvHe}$ and $\GT_\alpha$ when Wigner-SU(4) corrections are treated nonperturbatively.

\begin{figure}[hbt]
    \centering
    \includegraphics[width=0.75\linewidth]{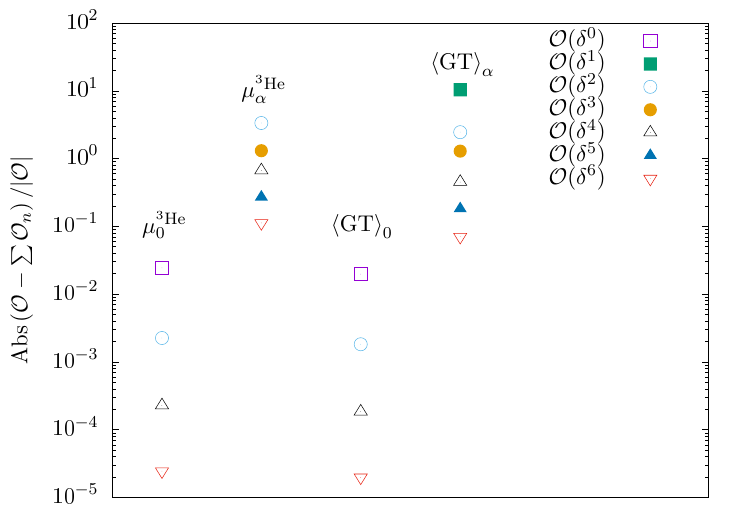}
    \caption{Plot of comparison between full physical values of $\GT_0$, $\mu_0^{\jjvHe}$, $\GT_{\alpha}$, and $\mu_{\alpha}^{\jjvHe}$ with no Wigner-SU(4) expansion and sum of these values up to a given order in $\delta$ expansion all the way up to $\mathcal{O}(\delta^6)$.}
    \label{fig:WigConvg}
\end{figure}

To further illustrate the Wigner-SU(4) expansion in $\delta$ for $\GT_0$, $\mu_0^{\jjvHe}$, $\GT_{\alpha}$, and $\mu_{\alpha}^{\jjvHe}$ we plot in Fig.~\ref{fig:WigConvg} the difference between their values, when Wigner-SU(4) is treated nonperturbatively, and the sum of their Wigner-SU(4) correction to $\mathcal{O}(\delta^n)$ all divided by their values when Wigner-SU(4) is treated nonperturbatively.
Again we see that $\mu_{0}^{\jjvHe}$ and $\GT_0$ appear to converge nicely to their values when Wigner-SU(4) is treated nonperturbatively.  The values of $\mu_{\alpha}^{\jjvHe}$ and $\GT_{\alpha}$ also appear to converge to their values when Wigner-SU(4) is treated nonperturbatively, however they start much further away and thus by $\mathcal{O}(\delta^6)$ are still noticeably far away.

\section{\label{sec:conclusion}Conclusions}

In this work we calculated the Coulomb corrections to $\mu^{\jjvHe}$ and the GT matrix element for $\jjvH$ $\beta$-decay in \EFT.  We found the Coulomb corrections to be far smaller than naively expected from \EFT power counting.  As in Ref.~\cite{Vanasse:2017kgh} for $\mu^{\jjvHe}$ at LO we find in units of nuclear magnetons -1.8675(6450), and up to and including NLO we find $-2.136(255)$, when fitting $L_1$ to the $\jjvH$ magnetic moment. This agrees well with the experimental value of -2.127.  The $\mathcal{O}(\alpha)$ Coulomb correction to $\mu^{\jjvHe}$ computed in this work is $0.00335(116)$, $\approx\!0.18\%$ of the LO value of -1.8675.  This is much smaller than the naive expected correction of $\alpha M_N/p^*\sim 8\%$. A recent calculation of $\mu^{\jjvHe}$, that treated Coulomb interactions nonperturbatively, found similarly sized Coulomb corrections~\cite{Lin:2026wln}.

We also calculated the GT matrix element for $\jjvH$ $\beta$-decay at LO in \EFT finding 0.9807(3387), as in Ref.~\cite{Nguyen:2024rlr}, and in this work the $\mathcal{O}(\alpha)$ correction for which we found $-0.000740(256)$, $\approx\!\!0.075\%$ of the LO value.   Again this is much smaller than the naive expected error of $\alpha M_N/p^*\!\!\sim\! 8\%$.  Calculating the GT matrix element to NLO we fit the LEC $l_{1,A}$ to the triton half life, while including Coulomb corrections.  Using this value of $l_{1,A}$ we then determined the reduced matrix element for $pp$ fusion of $\Lambda(p)=2.776(331)$.  This is in agreement with the phenomenological value of $\Lambda(0)=2.65(1)$~\cite{Adelberger:2010qa}.  Given the smallness of the $\mathcal{O}(\alpha)$ correction to $\GT$ our value for $\Lambda(0)$ is very close to the NLO \EFT calculation of Ref.~\cite{Nguyen:2024rlr}, where $\mathcal{O}(\alpha)$ Coulomb corrections were not included.

In order to explain the unnaturally small size of the $\mathcal{O}(\alpha)$ corrections to $\mu^{\jjvHe}$ and $\GT$ we transformed these corrections to the Wigner-SU(4) basis and performed a Wigner-SU(4) expansion in terms of the Wigner-SU(4) symmetry breaking parameter $\delta$~\cite{Vanasse:2016umz}.  Doing this expansion we found that the $\mathcal{O}(\alpha)$ Coulomb correction to the $\jjvHe$ magnetic moment was suppressed by $\mathcal{O}(\delta^2)$ while the $\mathcal{O}(\alpha)$ Coulomb correction to the $\GT$ matrix element of $\jjvH$ $\beta$-decay was suppressed by $\mathcal{O}(\delta)$.  Assuming $\mathcal{O}(\delta)\sim\mathcal{O}(\rho^2)$ we found a revised estimate for $\mu_{\alpha}^{\jjvHe}$ ($\GT_{\alpha}$) of $\rho^4\alpha M_N/p^*\!\!\sim\!0.11\%$ ($\rho^2\alpha M_N/p^*\!\!\sim\!0.92\%$).  While for $\mu_{\alpha}^{\jjvHe}$ this is quite close to the observed value of $0.18\%$ it is considerably larger than the $0.075\%$ for $\GT_{\alpha}$.  To understand this better we carried out an order-by-order expansion in $\delta$, up to $\mathcal{O}(\delta^6)$, for $\GT_{\alpha}$ and $\mu_{\alpha}^{\jjvHe}$ as well as $\GT_0$ and $\mu_{0}^{\jjvHe}$.  Carrying out this expansion we found that $\GT_0$ and $\mu_0^{\jjvHe}$ were reproduced well at $\mathcal{O}(\delta^6)$ while $\GT_{\alpha}$ and $\mu_{\alpha}^{\jjvHe}$ were not.  We found that the leading $\mathcal{O}(\delta)$ [$\mathcal{O}(\delta^2)$] contribution for $\GT_{\alpha}$ [$\mu_{\alpha}^{\jjvHe}$] was considerably larger than its value when Wigner-SU(4) is treated nonperturbatively.  However, we did observe that $\GT_{\alpha}$ and $\mu_{\alpha}^{\jjvHe}$ appear to be converging in the Wigner-SU(4) expansion but every other order in $\delta$ has the opposite sign and is not insignificant compared to the previous order of $\delta$.  This could simply be a sign of the breakdown of the Wigner-SU(4) expansion for the $\mathcal{O}(\alpha)$ correction or signs of some as of yet unknown expansion.

In the future we would like to better understand the observed pattern of convergence in the Wigner-SU(4) expansion.  We would also like to carry out the \EFT calculation to higher orders.  The calculation of  $\mu^{\jjvHe}$ is complicated by the fact that a three-nucleon magnetic current counterterm is necessary at NNLO~\cite{Lin:2022yaf}, which would need to be fit to some other three-body datum involving this current.  Based on this we also expect to see a similar counterterm for the axial current in the three-nucleon system.  Again meaning that a NNLO calculation of $\jjvH$ $\beta$-decay would require another three-nucleon observable involving the axial current.  It would also be interesting to see if the $\mathcal{O}(\alpha^2)$ corrections suffer the same fate of suppression as their $\mathcal{O}(\alpha)$ counterparts.

\acknowledgments{We thank Roxanne Springer and Xincheng Lin for useful discussions.  We also thank Daniel Phillips for useful feedback on the presentation of this paper at the APS DNP 2025 Fall Meeting.  Ha Nguyen is supported by the Henry W. Newson Fellowship and the U.S. Department of Energy, Office of Science, Office of Nuclear Physics, under Award Number DE-FG02-05ER41368.}

\appendix

\section{\label{app:chi}$\boldsymbol{\chi}$ Functions}

The $\boldsymbol{\chi_j}$ functions, where $j=a,b,c,d$, and $e$ appear in Eqs.~\eqref{eq:LOForm},\eqref{eq:NLOForm}, and~\eqref{eq:alphaForm}.  $\boldsymbol{\chi}_a$ is given by
\begin{align}
    &\boldsymbol{\chi}_a\left(E,\vect{K},\vect{P},p_0,k_0,\vect{p},\vect{k}\right)=ie(2\pi)^4\delta(k_0-p_0)\boldsymbol{\delta}^3\left(\vect{k}-\vect{p}-\frac{2}{3}\vect{Q}\right)\\\nonumber
    &i\Dbb\left(\frac{2}{3}E+k_0,\vect{k}+\frac{2}{3}\vect{K}\right)\frac{i}{\frac{1}{3}E-k_0-\frac{\left(\vect{k}-\frac{1}{3}\vect{K}\right)^2}{2M_N}+i\epsilon}\frac{i}{\frac{1}{3}E-k_0-\frac{\left(\vect{k}-\frac{2}{3}\vect{Q}-\frac{1}{3}\vect{P}\right)^2}{2M_N}+i\epsilon}\Yb_a,
\end{align}
where 
\begin{equation}
    \Dbb(E_B,p)=\Db\left(E_B+\frac{p^2}{2M_N},p\right).
\end{equation}
The 3$\times$3 c.c.~space matrix $\Yb_a$ for our form factors of interest is given in Tab.~\ref{tab:Machi}.
\begin{table}[hbt]
    \begin{tabular}{|c|c|c|c|c|c|c|c|c|c|}
        \hline
       &$\left[\Yb_a\right]_{11}$   & $\left[\Yb_a\right]_{12}$ & $\left[\Yb_a\right]_{13}$ & $\left[\Yb_a\right]_{21}$ & $\left[\Yb_a\right]_{22}$ & $\left[\Yb_a\right]_{23}$ & $\left[\Yb_a\right]_{31}$ & $\left[\Yb_a\right]_{32}$ & $\left[\Yb_a\right]_{33}$ \\\hline
       $F_C^{\jjvHe}(0)$ & 1 & 0 & 0 & 0 & 3 & 0 & 0 & 0 & 0\\\hline
       $F_M^{\jjvHe}(0)$ & $-\frac{1}{3}\kappa_p$ & 0 & 0 & 0 & $3\kappa_p$ & 0 & 0 & 0 & $\frac{3}{2}\kappa_n$\\\hline
       $F_W^{GT}(0)$ & $\frac{1}{3}$ & 0 & 0 & 0 & 3 & 0 & 0 & 0 & 0\\\hline
       $F_W^{F}(0)$ & 1 & 0 & 0 & 0 & -3 & 0 & 0 & 0 & 0\\\hline
    \end{tabular}
    \caption{\label{tab:Machi} Matrix elements of $\Yb_a$ for form factors of interest.}
\end{table}
$\boldsymbol{\chi}_b$ is given by
\begin{align}
    &\boldsymbol{\chi}_b\left(E,\vect{K},\vect{P},p_0,k_0,\vect{p},\vect{k}\right)=i\frac{2\pi e}{M_N}i\Dbb\left(\frac{2}{3}E+k_0,\vect{k}+\frac{2}{3}\vect{K}\right)\frac{i}{\frac{1}{3}E-k_0-\frac{\left(\vect{k}-\frac{1}{3}\vect{K}\right)^2}{2M_N}+i\epsilon}\\\nonumber
    &\times\frac{i}{\frac{1}{3}E-p_0-\frac{\left(\vect{p}-\frac{1}{3}\vect{P}\right)^2}{2M_N}+i\epsilon}\frac{i}{\frac{1}{3}E+k_0+p_0-\frac{\left(\vect{k}+\vect{p}-\frac{1}{3}\vect{Q}+\frac{1}{3}\vect{K}\right)^2}{2M_N}+i\epsilon}\\\nonumber
    &\times\frac{i}{\frac{1}{3}E+k_0+p_0-\frac{\left(\vect{k}+\vect{p}+\frac{1}{3}\vect{Q}+\frac{1}{3}\vect{P}\right)^2}{2M_N}+i\epsilon}\Yb_b i\Dbb\left(\frac{2}{3}E+p_0,\vect{p}+\frac{2}{3}\vect{P}\right),
\end{align}
where the 3$\times$3 c.c.~space matrix $\Yb_b$ for our form factors of interest is given in Tab.~\ref{tab:Mbchi}.
\begin{table}[hbt]
    \begin{tabular}{|c|c|c|c|c|c|c|c|c|c|}
        \hline
       &$\left[\Yb_b\right]_{11}$   & $\left[\Yb_b\right]_{12}$ & $\left[\Yb_b\right]_{13}$ & $\left[\Yb_b\right]_{21}$ & $\left[\Yb_b\right]_{22}$ & $\left[\Yb_b\right]_{23}$ & $\left[\Yb_b\right]_{31}$ & $\left[\Yb_b\right]_{32}$ & $\left[\Yb_b\right]_{33}$ \\\hline
       $F_C^{\jjvHe}(0)$ & 0 & 0 & 3 & 0 & 0 & -3 & 3 & -3 & 0\\\hline
       $F_M^{\jjvHe}(0)$ & $-\frac{5}{3}\kappa_p$ & $\kappa_n$ & $\kappa_p$ & $\kappa_n$ & $-3\kappa_n$ & $3\kappa_p$ & $\kappa_p$ & $3\kappa_p$ & 0 \\\hline
       $F_W^{GT}(0)$ & $-\frac{5}{3}$ & -1 & 0 & -1 & -3 & 0 & 0 & 0 & -3\\\hline
       $F_W^{F}(0)$ & 1 & 3 & 0 & 3 & -3 & 0 & 0 & 0 & -3\\\hline
    \end{tabular}
    \caption{\label{tab:Mbchi} Matrix elements of $\Yb_b$ for form factors of interest.}
\end{table}
$\boldsymbol{\chi}_c$ is given by
\begin{align}
    &\boldsymbol{\chi}_c\left(E,\vect{K},\vect{P},p_0,k_0,\vect{p},\vect{k}\right)=\\\nonumber
    &i\frac{eM_N}{Q}(2\pi)^4\delta(k_0-p_0)\boldsymbol{\delta}^3\left(\vect{p}-\vect{k}-\frac{1}{3}\vect{Q}\right)\frac{i}{\frac{1}{3}E-k_0-\frac{\left(\vect{k}-\frac{1}{3}\vect{K}\right)^2}{2M_N}+i\epsilon}\\\nonumber
    &\arctan\left(\frac{Q}{2\sqrt{\frac{1}{4}\left(\vect{k}+\frac{2}{3}\vect{K}\right)^2-\frac{2}{3}M_NE-M_Nk_0}+2\sqrt{\frac{1}{4}\left(\vect{k}+\vect{Q}+\frac{2}{3}\vect{K}\right)^2-\frac{2}{3}M_NE-M_Nk_0}}\right)\\\nonumber
    &\times i\Dbb\left(\frac{2}{3}E+k_0,\vect{k}+\frac{2}{3}\vect{K}\right)\Yb_ci\Dbb\left(\frac{2}{3}E+k_0,\vect{k}+\vect{Q}+\frac{2}{3}\vect{K}\right),
\end{align}
where the 3$\times$3 c.c.~space matrix $\Yb_c$ for our form factor factors of interest is given in Tab.~\ref{tab:Mcchi}.
\begin{table}
    \begin{tabular}{|c|c|c|c|c|c|c|c|c|c|}
        \hline
       &$\left[\Yb_c\right]_{11}$   & $\left[\Yb_c\right]_{12}$ & $\left[\Yb_c\right]_{13}$ & $\left[\Yb_c\right]_{21}$ & $\left[\Yb_c\right]_{22}$ & $\left[\Yb_c\right]_{23}$ & $\left[\Yb_c\right]_{31}$ & $\left[\Yb_c\right]_{32}$ & $\left[\Yb_c\right]_{33}$ \\\hline
       $F_C^{\jjvHe}(0)$ & 2 & 0 & 0 & 0 & 6 & 0 & 0 & 0 & 6\\\hline
       $F_M^{\jjvHe}(0)$ & $\frac{4}{3}(\kappa_p+\kappa_n)$ & $2(\kappa_p-\kappa_n)$ & 0 & $2(\kappa_p-\kappa_n)$ & 0 & 0 & 0 & 0 & 0\\\hline
       $F_W^{GT}(0)$ & 0 & 0 & -2 & 0 & 0 & 0 & -2 & 0 & 0\\\hline
       $F_W^{F}(0)$ & 0 & 0 & 0 & 0 & 0 & 6 & 0 & 6 & 0\\\hline
    \end{tabular}
    \caption{\label{tab:Mcchi} Matrix elements of $\Yb_c$ for form factors of interest.}
\end{table}
$\boldsymbol{\chi}_d$ is given by
\begin{align}
    &\boldsymbol{\chi}_d\left(E,\vect{K},\vect{P},p_0,k_0,\vect{p},\vect{k}\right)=\\\nonumber
    &ie(2\pi)^4\delta(k_0-p_0)\boldsymbol{\delta}^3\left(\vect{p}-\vect{k}-\frac{1}{3}\vect{Q}\right)\frac{i}{\frac{1}{3}E-k_0-\frac{\left(\vect{k}-\frac{1}{3}\vect{K}\right)^2}{2M_N}+i\epsilon}\\\nonumber
    &i\Dbb\left(\frac{2}{3}E+k_0,\vect{k}+\frac{2}{3}\vect{K}\right)\Yb_d i\Dbb\left(\frac{2}{3}E+k_0,\vect{k}+\vect{Q}+\frac{2}{3}\vect{K}\right),
\end{align}
where the 3$\times$3 c.c.~space matrix $\Yb_d$ for our form factor factors of interest is given in Tab.~\ref{tab:Mdchi}.
\begin{table}
    \begin{tabular}{|c|c|c|c|c|c|c|c|c|c|}
        \hline
       &$\left[\Yb_d\right]_{11}$   & $\left[\Yb_d\right]_{12}$ & $\left[\Yb_d\right]_{13}$ & $\left[\Yb_d\right]_{21}$ & $\left[\Yb_d\right]_{22}$ & $\left[\Yb_d\right]_{23}$ & $\left[\Yb_d\right]_{31}$ & $\left[\Yb_d\right]_{32}$ & $\left[\Yb_d\right]_{33}$ \\\hline
       $F_C^{\jjvHe}(0)$ & $-c_{0t}$ & 0 & 0 & 0 & $-3c_{0s}$ & 0 & 0 & 0 & $-3c_{0s}$\\\hline
       $F_M^{\jjvHe}(0)$ & $\frac{2}{3}M_NL_2$ & $\frac{1}{3}M_NL_1$ & 0 & $\frac{1}{3}M_NL_1$ & 0 & 0 & 0 & 0 & 0\\\hline
       $F_W^{GT}(0)$ & 0 & 0 & $-\frac{1}{g_A}l_{1,A}$ & 0 & 0 & 0 & $-\frac{1}{g_A}l_{1,A}$ & 0 & 0\\\hline
       $F_W^{F}(0)$ & 0 & 0 & 0 & 0 & 0 & $-3l_{1,V}$ & 0 & $-3l_{1,V}$ & 0\\\hline
    \end{tabular}
    \caption{\label{tab:Mdchi} Matrix elements of $\Yb_d$ for form factors of interest.}
\end{table}
Finally $\boldsymbol{\chi}_e$ is given by
\begin{align}
    &\boldsymbol{\chi}_e\left(E,\vect{K},\vect{P},p_0,k_0,\vect{p},\vect{k}\right)=-e(2\pi)^4\delta(k_0-p_0)\boldsymbol{\delta}^3\left(\vect{k}-\vect{p}-\frac{2}{3}\vect{Q}\right)\\\nonumber
    &i\mathbf{R}_1\left(\frac{2}{3}E+k_0+\frac{1}{2M_N}\left(\vect{k}+\frac{2}{3}\vect{K}\right)^2,\vect{k}+\frac{2}{3}\vect{K}\right)i\Dbb\left(\frac{2}{3}E+k_0,\vect{k}+\frac{2}{3}\vect{K}\right)\\\nonumber
    &\frac{i}{\frac{1}{3}E-k_0-\frac{\left(\vect{k}-\frac{1}{3}\vect{K}\right)^2}{2M_N}+i\epsilon}\frac{i}{\frac{1}{3}E-k_0-\frac{\left(\vect{k}-\frac{2}{3}\vect{Q}-\frac{1}{3}\vect{P}\right)^2}{2M_N}+i\epsilon}\Yb_a,
\end{align}

\section{\label{app:boostGalpha} Boosted $\Gb_\alpha$}
To calculate the boosted $\mathcal{O}(\alpha)$ Coulomb correction to the LO vertex function, Eq.~\eqref{eq:Galphaboost}, it is necessary to calculate the diagrams in Fig.~\ref{fig:Coulomb-inhom} in the boosted reference frame.  The kernel $\mathbf{K}_{\alpha}(q,k,E,\vect{K},k_0)$ from Eq.~\eqref{eq:Galphaboost} can be broken up into contributions from these individual diagrams yielding
\begin{align}
    \mathbf{K}_{\alpha}(q,k,E,\vect{K},k_0)=&\mathbf{K}_{\alpha}^{(V_1)}(q,k,E,\vect{K},k_0)+\mathbf{K}_{\alpha}^{(V_2)}(q,k,E,\vect{K},k_0)\\\nonumber
    &+\mathbf{K}_{\alpha}^{(B)}(q,k,E,\vect{K},k_0)+\mathbf{K}_{\alpha}^{(C)}(q,k,E,\vect{K},k_0).
\end{align}
The contribution from the $(V_1)$ diagram in Fig.~\ref{fig:Coulomb-inhom}, $\mathbf{K}_{\alpha}^{(V_1)}(q,k,E,\vect{K},k_0)$, is given by
\begin{align}
    &\mathbf{K}_{\alpha}^{(V_1)}(q,k,E,\vect{K},k_0)=4\pi M_N\alpha\frac{1}{qk}\\\nonumber
    &F_1\left[2\sqrt{\frac{3}{4}q^2-\frac{1}{2}k^2-\frac{2}{3}M_NE_B-M_Nk_0+\frac{1}{3}\vect{k}\cdot\vect{K}},2d(q),q,2k\right]\\\nonumber
    &\left(\begin{array}{rrr}
    0 & 0 & -3\\[-2mm]
    0 & 0 & 1\\[-2mm]
    0 & 0 & 0
    \end{array}\right)\Db\left(E_B,\vect{q}\right),
\end{align}
while the contribution of $(V_2)$ from Fig~\ref{fig:Coulomb-inhom}, $\mathbf{K}_{\alpha}^{(V_2)}(q,k,E,\vect{K},k_0)$, is
\begin{align}
    &\mathbf{K}_{\alpha}^{(V_2)}(q,k,E,\vect{K},k_0)=4\pi M_N\alpha\\\nonumber
    &\frac{1}{kq}F_1\left[2\sqrt{\frac{1}{4}k^2-\frac{2}{3}M_NE_B-M_Nk_0+\frac{1}{3}\vect{k}\cdot\vect{K}},2\sqrt{\frac{1}{4}k^2-\frac{2}{3}M_NE_B-M_Nk_0+\frac{1}{3}\vect{k}\cdot\vect{K}},k,2q\right]\\\nonumber
    &\left(\begin{array}{ccc}
    0 & 0 & 0\\[-2mm]
    0 & 0 & 0\\[-2mm]
    -2 & 2 & 0
    \end{array}\right)\Db\left(E_B,\vect{q}\right).
\end{align}
$\mathbf{K}_{\alpha}^{(B)}(q,k,E,\vect{K},k_0)$ comes from diagram ($B$) of Fig.~\ref{fig:Coulomb-inhom} and is given by
\begin{align}
    &\mathbf{K}_{\alpha}^{(B)}(q,k,E,\vect{K},k_0)=4\pi M_N\alpha\\\nonumber
    & \frac{1}{kq}F_1\left[0,2d(q)+2\sqrt{\frac{1}{4}k^2-\frac{2}{3}M_NE_B-M_Nk_0+\frac{1}{3}\vect{k}\cdot\vect{K}},q,k\right]\\\nonumber
   &\left(\begin{array}{ccc}
   1 & 0 & 0\\[-2mm]
   0 & 1 & 0 \\[-2mm]
   0 & 0 & 0
   \end{array}\right)\Db\left(E_B,\vect{q}\right).
\end{align}
Finally, $\mathbf{K}_{\alpha}^{(C)}(q,k,E,\vect{K},k_0)$ from diagram ($C$) in Fig.~\ref{fig:Coulomb-inhom} is given by
\begin{align}
    &\mathbf{K}_{\alpha}^{(C)}(q,k,E,\vect{K},k_0)=2\pi M_N\alpha\int\!\!d\Omega_q\mathcal{J}(\vect{q},\vect{k},\vect{K},k_0)\left(\begin{array}{rrr}
    1 & -3 & 0\\[-2mm]
    -1 & -1 & 0 \\[-2mm]
    0 & 0 & 0
    \end{array}\right)\Db\left(E_B,\vect{q}\right),
\end{align}
where
\begin{align}
    \mathcal{J}(\vect{q},\vect{k},\vect{K},k_0)&=\int\!\!\frac{d^3\ell}{(2\pi)^3}\\\nonumber
    &\frac{1}{\left(\vect{\ell}+\vect{q}+\frac{1}{2}\vect{k}\right)^2+\frac{1}{4}k^2-\frac{2}{3}M_NE_B-M_Nk_0+\frac{1}{3}\vect{k}\cdot\vect{K}-i\epsilon}\\\nonumber
    &\frac{1}{\left(\vect{\ell}+\vect{k}+\frac{1}{2}\vect{q}\right)^2+\frac{3}{4}q^2-M_NE_B-i\epsilon}\frac{1}{\ell^2}.
\end{align}
In principle the five dimensional integral for $\mathbf{K}_{\alpha}^{(C)}(q,k,E,\vect{K},k_0)$ can be reduced to an analytical form with a single dimensional integral.  However, in our calculation we take the derivative with respect to $k_0$, then set $\vect{K}=0$ and $k_0=\frac{1}{3}E_B-\frac{k^2}{2M_N}$ resulting in a significant simplification of this integral such that it gives the form of diagram (f) in Fig.~\ref{fig:alphaFF} but with different c.c.~space matrix elements.


%

\end{document}